\documentclass[reprint, prd, superscriptaddress, tightenlines, longbibliography, nofootinbib, eqsecnum, amsfonts, amsmath, floatfix, notitlepage]{revtex4-1}
\pdfoutput=1

\usepackage[utf8]{inputenc}
\usepackage{mathrsfs}
\usepackage{euscript}
\usepackage{graphics}
\usepackage{graphicx}
\usepackage{amsmath}
\usepackage{amssymb}
\usepackage{gensymb}
\usepackage{bm}
\usepackage[usenames,dvipsnames,svgnames,table]{xcolor}
\definecolor{linkcolor}{rgb}{0.6,0,0}
\definecolor{citecolor}{rgb}{0,0,0.75}
\definecolor{urlcolor}{rgb}{0.12,0.46,0.7}
\usepackage[breaklinks, colorlinks, urlcolor=urlcolor, linkcolor=linkcolor,citecolor=citecolor,pdfencoding=auto]{hyperref}
\usepackage{xspace}
\usepackage{wasysym}
\usepackage{times}
\usepackage{appendix}
\usepackage{comment}
\usepackage{lipsum}
\usepackage[nolist,nohyperlinks]{acronym}
\usepackage{float}
\usepackage{simplewick}
\usepackage{natbib, ifthen}
\usepackage{longtable}
\usepackage{mhchem} % for upper-left scripts
\usepackage[caption=false]{subfig}
\usepackage{multirow}
\usepackage{listings}

\newcommand{\la}{\langle}
\newcommand{\ra}{\rangle}

\newcommand{\mksym}[1]{\ifmmode {\rm #1}\else #1\fi}

\providecommand{\gea}{\ga}

\providecommand{\agt}{\gea}
\providecommand{\text}[1]{\rm{#1}}

\newcommand{\tot}{{\text{tot}}}

\providecommand{\CAMB}{{\tt camb}}

\newcommand{\begm}{\begin{pmatrix}}
\newcommand{\enm}{\end{pmatrix}}

\newcommand\ba{\begin{eqnarray}}
\newcommand\ea{\end{eqnarray}}
\newcommand\bea{\begin{eqnarray}}
\newcommand\eea{\end{eqnarray}}

\newcommand\be{\begin{equation}}
\newcommand\ee{\end{equation}}

\newcommand{\vphi}{{\boldsymbol{\phi}}}

\providecommand{\var}{\text{var}}
\providecommand{\Tr}{\text{Tr}}
\newcommand{\mC}{\bm{C}}

\newcommand{\mM}{\bm{M}}
\newcommand{\mN}{\bm{N}}

\newcommand{\mW}{\bm{W}}

\newcommand{\boldvec}[1]{{\mbox{\boldmath{$#1$}}}}

\newcommand{\vf}{\boldvec{f}}
\newcommand{\vg}{\boldvec{g}}

\newcommand{\clf}{\mathcal{F}}

\newcommand{\beq}{\begin{equation}}
\newcommand{\enq}{\end{equation}}
\newcommand{\beqa}{\begin{eqnarray}}
\newcommand{\enqa}{\end{eqnarray}}
\newcommand{\beit}{\begin{itemize}}
\newcommand{\enit}{\end{itemize}}
\newcommand{\bem}{\begin{pmatrix}}
\renewcommand{\enm}{\end{pmatrix}}
\newcommand{\vecL}{\mathbf L}

\newcommand{\vecalpha}{\boldsymbol{\alpha}}	

\newcommand{\vecx}{\mathbf{x} }
\newcommand{\vecy}{\mathbf{y} }

\renewcommand{\Tr}{\mathrm{Tr}}
\newcommand{\Var}[1]{\textrm{Var}\left(#1\right)}

\newcommand{\obs}{\textrm{obs}}

\newcommand{\eff}{\textrm{eff}}
\providecommand{\fsky}{f_{\mathrm{sky}}}
\newcommand{\fpatch}{f_{A,L}}
\renewcommand{\max}{\mathrm{max}}
\renewcommand{\min}{\mathrm{min}}

\renewcommand{\bem}{\begin{bmatrix}}
\renewcommand{\enm}{\end{bmatrix}}
\newcommand{\vecn}{\mathbf n}

\newcommand{\F}{ {\boldsymbol F}}

\newcommand{\N}{ {\boldsymbol N}}

\newcommand{\vecell}{ {\boldsymbol  \ell}}

\newcommand{\FWHM}{{\rm FWHM}}

\newcommand{\lensit}{\textsc{LensIt}\xspace}

\newcommand{\g}{g}

\newcommand{\MC}{\text{MC}}

\newcommand{\RDN}{\ce{^{\text{(RD)}}N}}
\newcommand{\MCN}{\ce{^{\text{(MC)}}N}}
\newcommand{\fid}{\rm fid}

\newcommand{\mb}{\bm{b}}
\newcommand{\noise}{\rm noise}
\newcommand{\hits}{\rm hits}
\newcommand{\filt}{\rm filt}

\newcommand{\vnabla}{\boldsymbol{\nabla}}	
{ % padding tables
\newcommand{\MF}{\rm MF}
\newcommand{\s}{s}

\newcommand{\one}{\textbf{1}}
\newcommand{\two}{\textbf{2}}
\newcommand{\three}{\textbf{3}}
\newcommand{\four}{\textbf{4}}
\newcommand{\five}{\textbf{5}}
\newcommand{\R}{\mathcal{R}}
\newcommand{\vex}{\boldsymbol{X}}

\newcommand{\cvec}{\boldsymbol{C}}

\let\oldsqrt\sqrt
\def\sqrt{\mathpalette\DHLhksqrt}
\def\DHLhksqrt#1#2{%
\setbox0=\hbox{$#1\oldsqrt{#2\,}$}\dimen0=\ht0
\advance\dimen0-0.2\ht0
\setbox2=\hbox{\vrule height\ht0 depth -\dimen0}%
{\box0\lower0.4pt\box2}}

\usepackage[dvipsnames]{xcolor}

\tolerance=1
\emergencystretch=\maxdimen
\hyphenpenalty=10000
\hbadness=10000
\begin{document}
\newcommand{\Sussex}{Department of Physics \& Astronomy, University of Sussex, Brighton BN1 9QH, UK}
\author{Mark Mirmelstein}
\affiliation{\Sussex}
\author{Julien Carron}
\affiliation{\Sussex}
\author{Antony Lewis}
\affiliation{\Sussex}

%\preprint{}

\title{Optimal filtering for CMB lensing reconstruction}

% address or URL should go in the {}'s for \email and \homepage.
% Please use the appropriate macro for each type of information

%\author{Authors}
%\affiliation{Department of Physics and Astronomy, University of Sussex, Brighton BN1 9QH, UK}

%\thanks{}


\date{\today}
\begin{abstract}
Upcoming ground-based cosmic microwave background experiments will provide CMB maps with high sensitivity and resolution that can be used for high fidelity lensing reconstruction. However, the sky coverage will be incomplete and the noise highly anisotropic, so optimized estimators are required to extract the most information from the maps. We focus on quadratic-estimator based lensing reconstruction methods that are fast to implement, and compare new more-optimally filtered estimators with various estimators that have previously been used in the literature. Input CMB maps can be optimally inverse-signal-plus-noise filtered using conjugate gradient (or other) techniques to account for the noise anisotropy. However, lensing reconstructions from these filtered input maps have an anisotropic response to the lensing signal and are difficult to interpret directly. We describe a second-stage filtering of the lensing maps and analytic response model that can be used to construct lensing power spectrum estimates that account for the anisotropic response and noise inhomogeneity in an approximately optimal way while remaining fast to compute. We compare results for simulations of upcoming Simons Observatory and CMB Stage-4 experiments to show the robustness of the more optimal lensing reconstruction pipeline and quantify the improvement compared to less optimal estimators. We find a substantial improvement in reconstructed lensing power variance between optimal anisotropic and isotropic filtering of CMB maps, and up to 30\% improvement in variance by using the additional filtering step on the reconstruction potential map. Our approximate analytic response model is unbiased to within a small percent-level additional Monte Carlo correction.
\end{abstract}

%\pacs{98.80.-k 98.65.Dx}
%\keywords{}

\maketitle

%%%%%%%%%%%%%%%%%%%%%%%%%%%%%%%%%%%%%%%%%%%%
%%%%%%%%%%%%%%%%%%%%%%%%%%%%%%%%%%%%%%%%%%%%
\section{Introduction}
\label{sec:intro}

Gravitational lensing of the cosmic microwave background (CMB) can be measured from the small changes induced in the observed temperature and polarization anisotropies (see Ref.~\cite{Lewis:2006fu} for a review).
Precision observations of the CMB can therefore be used to reconstruct the lensing potential, and constrain the evolution and geometry of the Universe between recombination and today. Planck has measured the lensing signal over $70\,\%$ of the sky~\cite{PL2018}, but the lensing reconstruction remains noise-dominated on most scales due to the limited resolution and sensitivity of the CMB measurements. In the upcoming years, new ground-based CMB experiments will substantially improve current measurements, with Simons Observatory (SO)~\cite{Ade:2018sbj} and then CMB-S4~\cite{Abazajian:2016yjj} (hereafter S4) building on the ongoing observations by ACTpol~\cite{Sherwin:2016tyf}, SPTpol~\cite{Story:2014hni} and POLARBEAR~\cite{Ade:2017uvt}. However, the instrumental noise affecting the B-mode polarization used for lensing reconstruction will remain important, and the lensing reconstruction noise from the instrumental noise and cosmic variance of the unlensed CMB will continue to dominate for small-scale lensing reconstruction modes ($L\agt 400$ for SO)\footnote{We follow the standard convention and use $L$ rather than $\ell$ for lensing multipoles.}.
The instrumental noise also typically varies substantially over the observed sky area due to the exact way in which the sky is repeatedly scanned. It is therefore important to consider how to account for the inhomogeneous noise properties to exploit the CMB maps in an efficient way.

Since the noise and unlensed CMB are expected to be Gaussian, and lensing simply deflects points on the sky, it is straightforward to write down a likelihood for the observed CMB given a fixed lensing deflection field. The lensing potential is also Gaussian to a good approximation on most relevant scales, so finding the lensing potential that maximizes the posterior then gives an optimal estimator for the lensing potential~\cite{Hirata:2003ka,Carron:2017mqf}. The resulting estimator is a complicated non-linear function of observed fields that has to be evaluated iteratively, but can be approximated to good accuracy for the near future by a quadratic estimator (QE) that is easier to evaluate~\cite{Hirata:2002jy,Okamoto:2003zw,Hanson:2009kr}.
To be optimal with anisotropic noise, the QE must be written as a quadratic function of filtered observed CMB fields, where the inverse-signal-plus-noise filters act to down-weight areas of the sky with higher noise (or foreground power). In this paper we compare this optimal filtering with various simpler filtering methods that have been used in the literature, to assess the improvement that can be obtained by using optimal filtering with upcoming experiments. We also present a new more optimal estimator that uses a second filtering step applied to the quadratic estimator reconstruction map.

Converting the lensing reconstruction map into an unbiased estimate of the lensing power spectrum (and any relevant cross-spectra) is important for parameter estimation, but is more complicated if the input maps are inhomogeneously filtered as the lensing reconstruction maps then effectively have a position-dependent normalization. This can be accounted for by applying a brute-force Monte Carlo (MC) correction, but we show that a simple analytic `patch' approximation is sufficient to capture the dominant effect. This approximation relies on the fact that the lensing estimators are quasi-local, so the lensing reconstruction at a given point mostly depends on the surrounding area of the observed CMB. If the noise varies slowly over the sky, we can therefore model the full signal as being made up of a set of patches within which the noise is nearly constant and the response can be calculated analytically. Corrections to this approximation can be measured by MC simulations and are perturbatively small, so the dominant signal can still be related analytically to the theoretical model.

Since the instrumental noise is inhomogeneous, the lensing reconstruction noise is also inhomogeneous, and an optimal spectrum estimator should account for this. It is possible to write down an approximate full likelihood for the lensing power spectrum and maximize it~\cite{Hirata:2003ka}; however, this again has to be solved iteratively, and makes the final lensing spectrum estimate a highly non-trivial function of the theoretical model parameters. Instead, we consider a simpler approximate solution that estimates the power spectrum from filtered QE maps, and assess the improvement using an approximate model for the reconstruction noise. The resulting power spectrum depends only on the four-point function of the CMB maps, and hence can be modelled straightforwardly as a function of parameters in a similar way to other QE-based analyses.

We start in Sec.~\ref{sec:methodology} by outlining the lensing reconstruction methodology and steps required to estimate the lensing power from the observed maps, and discuss various possible ways of filtering the CMB maps. In Sec.~\ref{sec:analytic} we develop the analytic patch approximation and assess its accuracy by comparison with simulations. Sec.~\ref{sec:results} presents our main results, where we show the improvement in signal-to-noise that can be obtained using the two optimal filtering steps. Throughout we assume a Gaussian unlensed CMB and inhomogeneous but uncorrelated pixel instrumental noise. For illustration we show results assuming a fiducial~$\Lambda$CDM model using parameters similar to those estimated by Planck~\cite{Ade:2015xua}.

%%%%%%%%%%%%%%%%%%%%%%%%%%%%%%%%%%%%%%%%%%%%
%%%%%%%%%%%%%%%%%%%%%%%%%%%%%%%%%%%%%%%%%%%%
\section{Methodology}
\label{sec:methodology}

We broadly follow the methodology of the Planck lensing analysis~\cite{PL2018}. For simplicity we use the flat-sky approximation when showing numerical results, using lensing analysis tools and simulated lensed CMB from \lensit\footnote{\href{https://github.com/carronj/LensIt}{https://github.com/carronj/LensIt}}.
Our lensing analysis is based on QEs, which can be evaluated efficiently in the case of full sky observations with isotropic noise using methods explained in Refs.~\cite{Okamoto:2003zw,Lewis:2006fu,Hanson:2009kr}. Here, we will briefly review the motivation for using quadratic estimates and explain the need for filtered maps.

Following Ref.~\cite{Hanson:2009kr}, we take the observed maps to be a vector of fields\footnote{Throughout, bold symbols are used to describe vectors or matrices.} $\vex$ (with temperature $T$, and/or polarization Stokes parameters $Q$ or $U$) with $\vex(\vecx)=\tilde{\vex}(\vecx)+{\bf n}(\vecx)$ in pixel space at position $\vecx$, where $\vecn$ is the instrumental noise realization which is assumed to be Gaussian, anisotropic and spatially uncorrelated. Here $\tilde{\vex}$ is the beamed and lensed CMB given in terms of the unlensed fields $\vex$ and deflection angle $\vecalpha$ by $\tilde{\vex}(\vecx) = \mb \vex(\vecx + \vecalpha)$ where $\mb$ accounts for the beam transfer function.
For a fixed lens model, we can write down the log-likelihood $\mathscr{L}$ of the observed CMB given the lensing potential $\phi$ (which is related to the deflection angle, $\vecalpha = \vnabla\phi$) as
\beq
-\mathscr{L}(\vex|\phi) = \frac{1}{2}{\vex}^\top \left(\mC_{\phi}^{\vex\vex}\right)^{-1}\vex+\frac{1}{2}\ln\left| \mC_{\phi}^{\vex\vex}\right|,
\enq
where $\mC_{\phi}^{\vex\vex} \equiv \mC_{\phi}^{\vex\vex}(\vecx,\vecy)=\mC_\phi^{\tilde{\vex}\tilde{\vex}}(\vecx,\vecy)+\mN(\vecx,\vecy)$ is the covariance of the map for fixed lensing potential $\phi$. Here the coordinates are integrated over in the contractions and $\mN$ is the covariance of the noise, which we take to be diagonal in pixel space and uncorrelated between $T, Q, U$. We want to maximize the likelihood with respect to $\phi(\vecx)$. This is done by equating the log-likelihood's derivative with respect to $\phi(\vecx)$ to zero,
\beq
\begin{split}
\frac{\delta \mathscr{L}}{\delta \phi(\vecx)} =&\; \frac{1}{2}{\vex}^\top \left(\mC_{\phi}^{\vex\vex}\right)^{-1}\frac{\delta \mC_{\phi}^{\vex\vex}}{\delta \phi(\vecx)}\left(\mC_{\phi}^{\vex\vex}\right)^{-1}\vex \\
&\; -\frac{1}{2}\Tr\left[\left( \mC_{\phi}^{\vex\vex}\right)^{-1}\frac{\delta \mC_{\phi}^{\vex\vex}}{\delta \phi(\vecx)}\right] \\
=&\; 0.
\end{split}
\enq
The trace term can be written as a `mean field' (MF) average of the first term ($\equiv \hat{\g}^{\phi}(\vecx)$, see Ref.~\cite{Hirata:2002jy,Hanson:2009kr}), so that we require a solution to $\hat{\g}^{\phi}(\vecx) - \la\hat{\g}^{\phi}(\vecx)\ra = 0$.
The general solution for $\phi(\vecx)$ has to be obtained iteratively, but the first step of Newton-Raphson from zero (denoted by a subscript of 0 below) gives the approximate QE solution
\beq
\hat \vphi = \bm{\clf}^{-1}\left(\hat{\vg}_0^{\phi} - \la\hat{\vg}_0^{\phi}\ra\right),
\label{eq:quad_est}
\enq
where $\hat{\vg}_0^{\phi}$ now depends on the CMB maps only via the inverse-variance filtered combination $\left(\mC_{\phi=0}^{\vex\vex}\right)^{-1}\vex$.
Here technically the covariance is evaluated with $\phi=0$, however the accuracy of the estimator can be improved beyond leading order by using $\left(\mC^{\vex\vex}\right)^{-1} \vex$, i.e. filtering using the covariance evaluated using lensed CMB power spectra~\cite{Hanson:2010rp,Lewis:2011fk}.

In Eq.~\eqref{eq:quad_est} the Hessian matrix of second derivatives of the log-likelihood is approximated by its expectation, the Fisher matrix $\bm{\clf}$. This normalizes the result, and can be evaluated analytically in the case of full sky and isotropic noise where it is diagonal in harmonic space; more generally it is non-trivial to evaluate exactly. The approximate optimal maximum likelihood solution involves inverse-variance filtering, but it is also possible to make other choices: as long as $\bm{\clf}$ is adjusted appropriately, the estimator can remain unbiased (at the expense of some increase in reconstruction noise compared to the optimal case).

In the limit of a perturbatively Gaussian lensing field, the optimal maximum-likelihood estimator for the power spectrum can also be derived from the likelihood function. As shown by Appendix B of Ref.~\cite{Schmittfull:2013uea} this reduces to a (bias-subtracted) QE up to a Fisher-normalization, where the QE is calculated using inverse-variance filtered fields. The CMB maps only enters via their inverse-variance filtered version, as for the estimator of $\phi$ itself, so our first filtering step of the CMB maps should remain perturbatively optimal. The optimal power spectrum estimator also only depends on the naive power spectrum of the (unnormalized) $\phi$ estimator, corresponding to the inverse-noise weighted $\phi$ estimator (in agreement with Ref.~\cite{PL2018} that for noise-dominated constructions uniform weighting of the inverse-noise filtered field is optimal). The limit of small $C_L^{\phi\phi}$ does not of course apply directly in the case of lensing reconstruction, where the signal can be measured at high signal-to-noise, but we can expect that for larger $C_L^{\phi\phi}$ the inverse-noise weighting will approximately generalize to inverse-signal-plus-noise weighting, which is consistent with the perturbatively-optimal estimator in the small-signal limit.

In general, the QE lensing power spectrum analysis consists of the following five stages:
\begin{enumerate}
\item[\bf A.]\textbf{Filtering the observed CMB maps}. As mentioned, there are several filtering methods one could use on the CMB maps before obtaining the QE. Here we compare the optimal method to alternative methods that have been used in the literature.

\item[\bf B.]\textbf{Constructing the quadratic lensing estimators}. The filtered fields are efficiently combined in real space to obtain the unnormalized lensing QE.

\item[\bf C.]\textbf{Subtracting the mean field}. The map-level MF signal expected from mask, noise and other anisotropic features of the map in the absence of lensing is subtracted. Here we also apply an approximate analytic and isotropic normalization at the map level, though this has no impact on the resulted lensing power as long as everything is done consistently.

\item[\bf D.]\textbf{Filtering the reconstructed $\boldsymbol\kappa$ map}.
This optional step is new to our analysis. Approximating the unnormalized convergence ($\kappa$) reconstruction map as depending locally on the CMB maps and their filtered versions of stage A, we can apply an approximate local normalization to obtain the noisy reconstructed $\kappa$ map and then inverse-variance filter it using an approximate model for the local variance.

\item[\bf E.]\textbf{Estimating the power spectrum}. The power spectrum of the filtered $\kappa$ is biased, due to reconstruction noise, other lensing contractions, masking and other non-idealities, and because it is not yet normalized (the unfiltered-$\kappa$ spectrum is also biased, but with different debiasing terms). We subtract an MC estimate of the $N_{0,L}$ reconstruction noise and $N_{1,L}$ signal contractions, both obtained using 500 MC simulations, and multiply the resulting spectrum by an analytic correction made using a patch approximation (the latter is needed as the normalization of the QE applied in Stage C is only a local approximation; even without filtering the $\kappa$ map, any analytic response model would need to be corrected due to the inhomogeneity of the noise in the map).
\end{enumerate}

\begin{figure}[!h]
\centering
\includegraphics[width=\columnwidth]{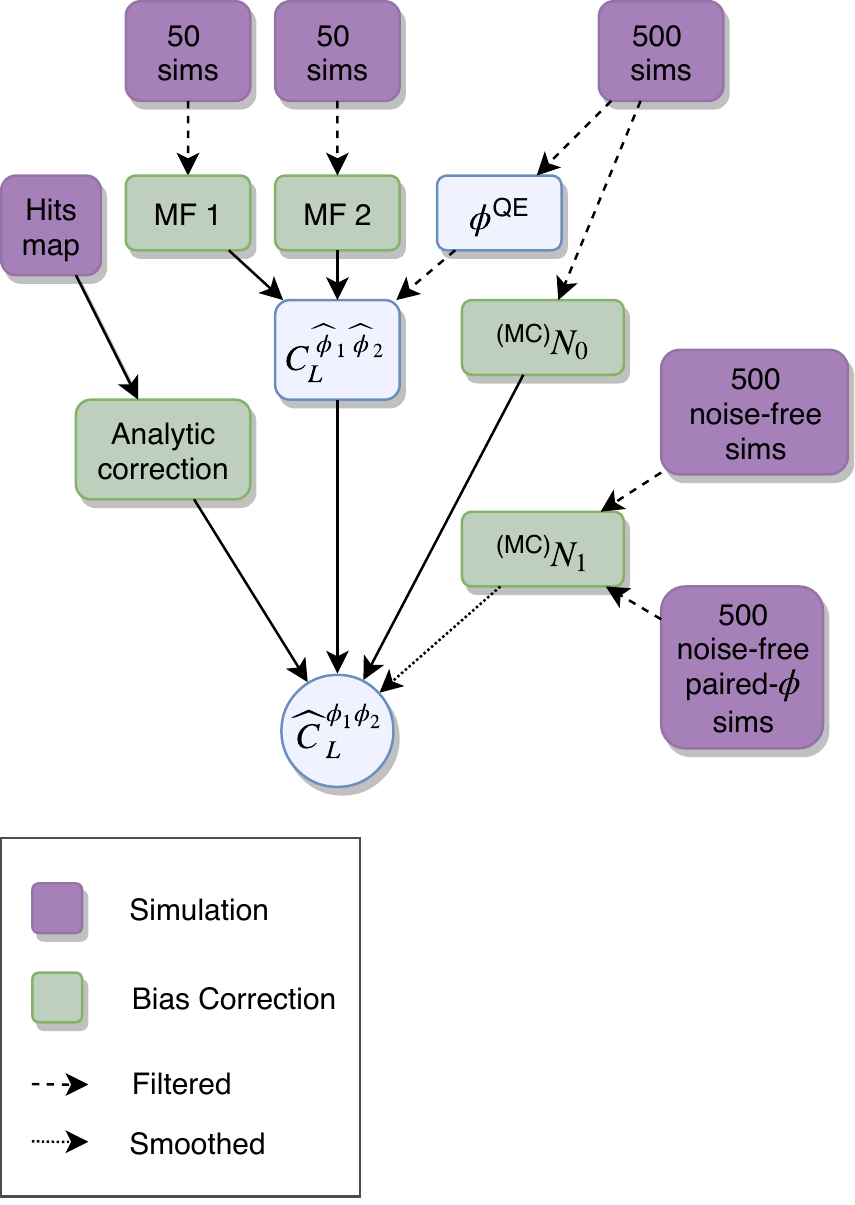}
\caption{The structure of the optimized QE lensing reconstruction pipeline which was used in this work. Here we use the same CMB realizations for constructing the QEs, $\MCN_{0,L}$ and $\MCN_{1,L}$ terms. The analytic correction is calculated using the simulated hit count map shown in Fig.~\ref{fig:weights}.}
\label{fig:methodology_flowchart}
\end{figure}

These steps are shown in Fig.~\ref{fig:methodology_flowchart} and discussed in more detail below.

%%%%%%%%%%%%%%%%%%%%%%%%%%%%%%%%%%%%%%%%%%%%
\subsection{Filtering methods}
\label{subsec:filtering_methods}

The first filtering step is a linear operation that is applied to the CMB maps giving a filtered CMB field
\begin{equation}
\bar \vex \equiv \mM \vex,
\end{equation}
where $\mM$ is some linear filtering matrix that's designed to removed masked areas and (optionally) down-weight noisier pixels or other noisy or contaminated modes. For optimal filtering $\mM$ is non-diagonal, and performs the full anisotropic inverse-variance filtering. For comparison we also consider approximate filtering methods where $\mM$ is taken to be diagonal in pixel space, which is somewhat faster to implement but less optimal.

\begin{figure}[!h]
\centering
\includegraphics[width=\columnwidth]{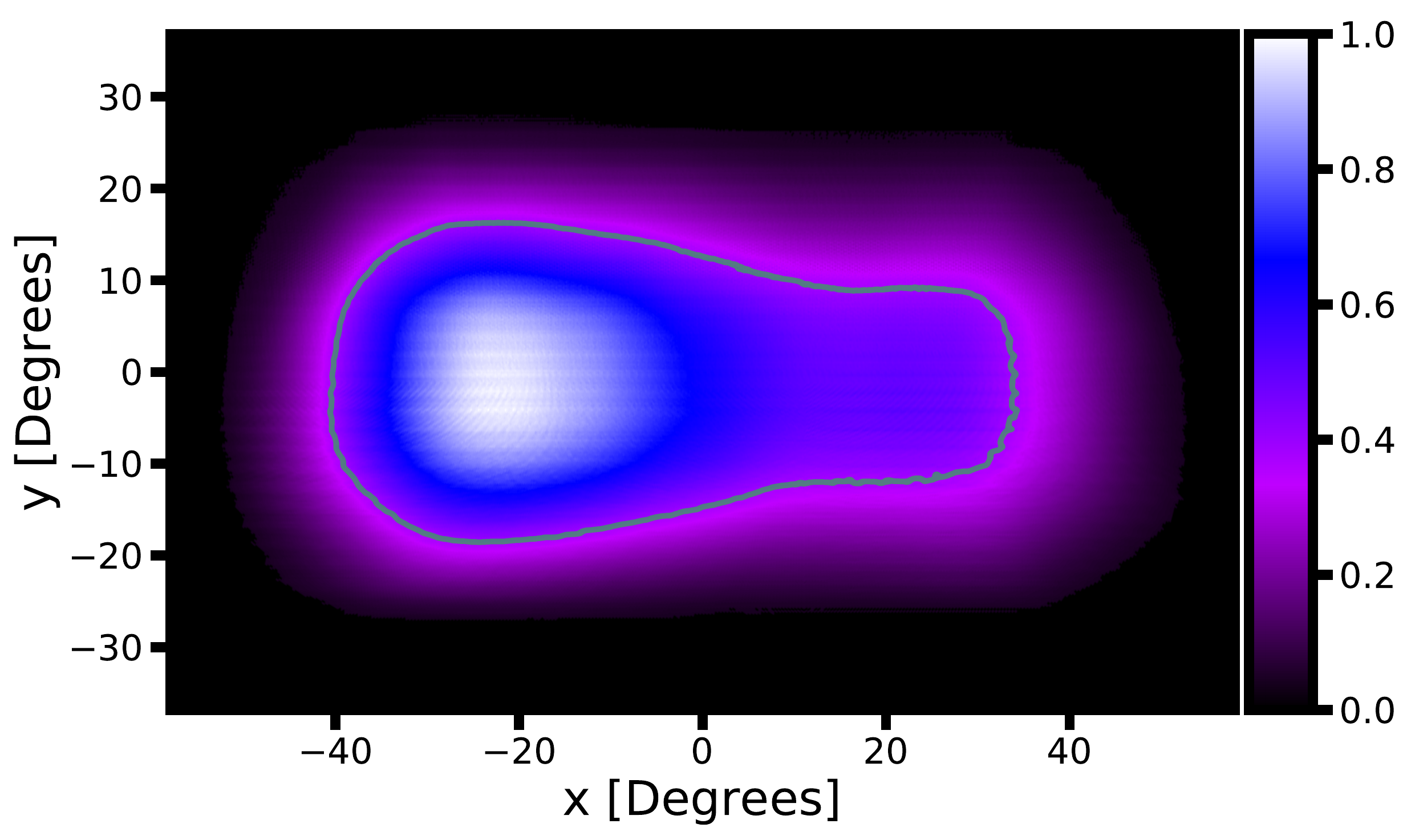}
\caption{The normalized hit count map used to simulate anisotropic noise in the simulations. The white pixels are those which were scanned for longer time, while the dark blue-purple fade shows pixels which were scanned less frequently. The black pixels are those which were not scanned. The area within the green boundary shows the unmasked regions used for the comparison isotropic-filtering analysis described in Sec.~\ref{sec:isotropic_filtering_on_masked_maps}. The normalized hit count map is also used as the weights for the comparison weighted maps analysis (Sec.~\ref{sec:isotropic_filtering_on_weighted_maps}).
The area with non-zero hits corresponds to 39\% of the flat-sky map we simulate, and about 13\% of the full sky area $4\pi$. The figure above is centred on the main observed area and does not show additional unobserved regions which all together form a 4096x4096 pixels map, which is 116 degrees on the side.
}
\label{fig:weights}
\vspace*{\floatsep}
\centering
\includegraphics[width=\columnwidth]{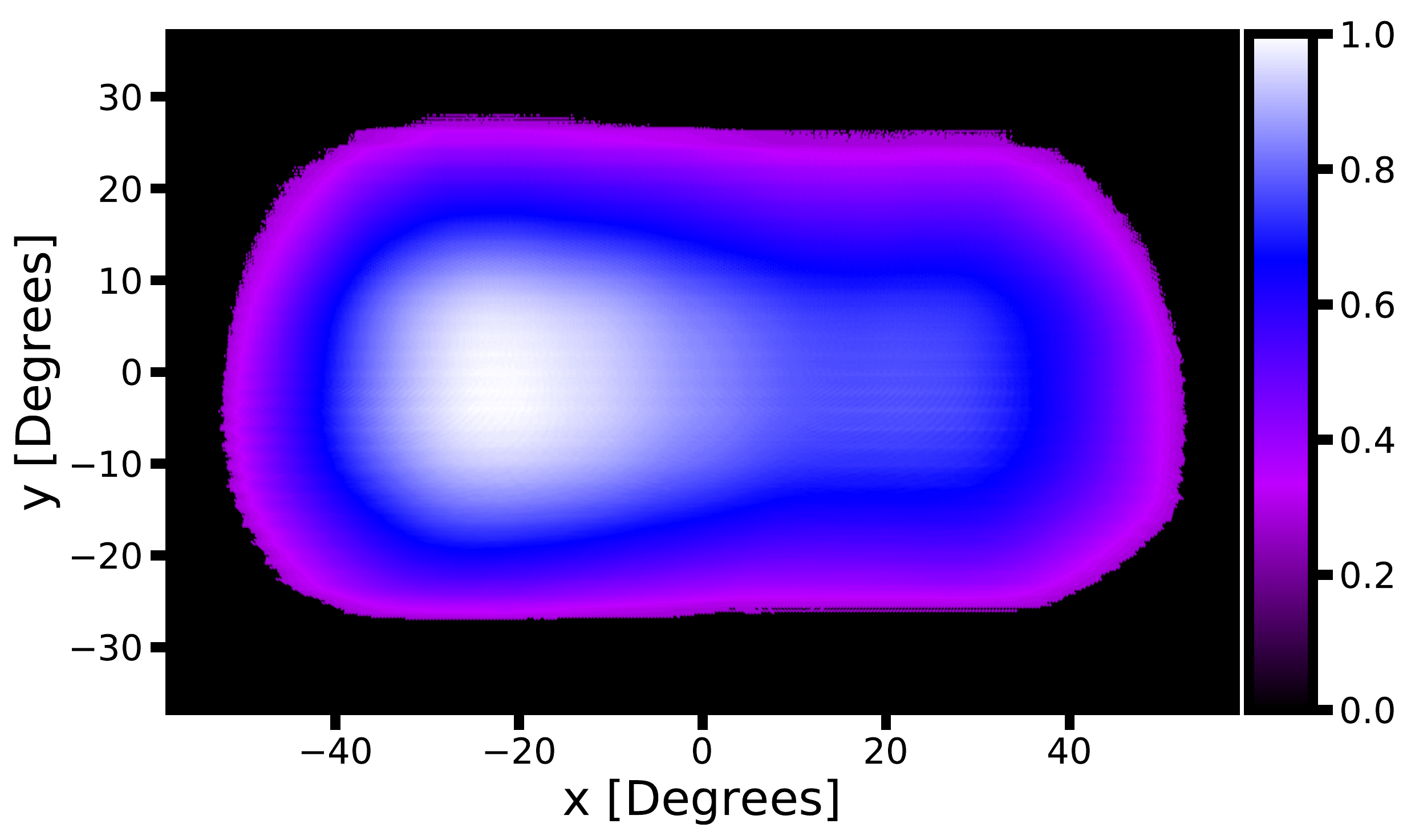}
\caption{The normalized inverse approximate lensing reconstruction noise $\left(N_{0,\eff}^{\kappa}(\vecx)\right)^{-1}$ from an analysis for an SO-like experiment using temperature and polarization. Compared to the hit count map in Fig.~\ref{fig:weights}, the reconstruction noise varies less strongly across the patch because it depends on the (isotropic) CMB variance as well as the (anisotropic) noise. The approximate effective white reconstruction noise $N_{0,\eff}^{\kappa}(\vecx)$ is defined as the average value of the isotropic $N_{0,L}^{\kappa}$ over the multipole range $40\leq L\leq 90$ on local approximately constant noise patches.
}
\label{fig:inv_Neff_map}
\end{figure}

We compare three different filtering methods by using them for lensing reconstruction on simulated sky maps with inhomogeneous noise. To simulate realistic noise inhomogeneity we use a part of a suggested scanning strategy for Simons Observatory's Small Aperture Telescope (SAT), the ``opportunistic'' scanning strategy presented in Ref.~\cite{Stevens:2018biw} as the hit count map. Note that while the lensing analysis of SO will come mainly from its Large Aperture Telescope (LAT), we chose to work with a part of the SAT scan as this scan has a well-contained area and is therefore more convenient for our flat-sky analysis.
To obtain the flat scanned area, we used the Cartesian projection from the HEALPix\footnote{\href{http://healpix.sf.net/}{http://healpix.sf.net/}} package to project the curved sky hit count map onto the flat sky.
The normalized hit count map $m_{\hits}(\vecx)$ is shown in Fig.~\ref{fig:weights}. The white pixels are seen more frequently during the sky scan simulation and so have a higher hit count (corresponding to lower noise), while the blue and purple areas show pixels which were scanned less often. Black pixels are not scanned.

We then use the hit count map to model the white noise standard deviation map
\beq
m_{\noise}(\vecx) = \sqrt{\frac{\s^2 N_{\hits,\tot}}{t_{\obs} m_{\hits}(\vecx)}},
\label{eq:noise_map}
\enq
where $\s$ is the instrument sensitivity for temperature or polarization, $t_{\obs}$ is the total observation time on the patch (with 1/5 efficiency), and $N_{\hits,\tot}$ is the total number of hits in the hit count map $m_{\hits}(\vecx)$. Using this model, noise is taken to be infinite outside the scanned region. The specifications we consider are discussed in Sec.~\ref{sec:results}.

In the following, each filtering method is presented along with motivation and assumptions. We filter modes below 4096 in the input maps using each method, and cut the filtered fields to $40\le\ell\le 3000$ before using them to obtain the QEs in stage B.

%%%%%%%%%%%%%%%%%%%%%%%%%%%%%%%%%%%%%%%%%%%%
\subsubsection{Optimal filtering}
\label{sec:optimal_filtering}

The inverse-variance filtered CMB maps can be written as
\begin{multline}
\bar \vex \equiv \left(\mb \cvec^{\fid} \mb^\top + \N\right)^{-1} \vex \\
=\left(\cvec^{\fid}\right)^{-1}\left[\left(\cvec^{\fid}\right)^{-1}+\mb^\top \N^{-1}\mb\right]^{-1}\mb^\top \N^{-1}\vex,
\label{eq:optfilt}
\end{multline}
where $\mb$ is the transfer function (here we consider a simple isotropic Gaussian beam with full-width half-maximum $\theta_{\FWHM}$) and $\cvec^{\fid}$ is a set of fiducial lensed power spectra. The noise can include variance due to foreground residuals, and a mask can be accounted for simply by taking the elements of $\N^{-1}$ to be zero in the corresponding pixels.

There are various possible ways to solve Eq.~\eqref{eq:optfilt}. We follow the multi-grid-preconditioned conjugate gradient method first demonstrated in the context of lensing by Ref.~\cite{Smith:2007rg} and then by the Planck and SPTpol lensing analyses~\cite{Ade:2013tyw,Story:2014hni,Ade:2015zua,PL2018}.
The pixel noise is in general spatially varying, but for simplicity the main
Planck lensing analyses approximate the noise as isotropic and white. Ref.~\cite{PL2018}
demonstrated that accounting for inhomogeneous noise in the filter can substantially improve Planck's noise-dominated polarization reconstruction, and in this paper we only use `optimal' to refer to the full anisotropic filter.

Rearranging~\eqref{eq:optfilt} to the form $\left(\mb C^{\fid} \mb^\top + \N\right) \bar \vex = \vex$, we define our stopping criterion for convergence to be when the norm of the difference of the two sides of the equation is smaller than $\epsilon|\vex|$.
We use $\epsilon = 10^{-5}$ for reconstructions using temperature and $\epsilon = 10^{-4}$ when using only polarization. We discuss some details of the choice of preconditioner and numerical performance in Appendix~\ref{app:cgconvergence}.

%%%%%%%%%%%%%%%%%%%%%%%%%%%%%%%%%%%%%%%%%%%%
\subsubsection{Isotropic filtering of masked maps}
\label{sec:isotropic_filtering_on_masked_maps}

The covariance matrix $\left(\mb C^{\fid} \mb^\top + \N\right)$ in Eq.~\eqref{eq:optfilt} is trivial to invert in harmonic space if the noise (and beam) is taken as isotropic, as both the theory and noise covariance matrices are then diagonal in harmonic space. We refer to this approximation as `isotropic filtering', which is substantially faster than optimal filtering since it avoids the conjugate-gradient inversion.
The isotropic filter is expected to be close to optimal over scales on which the corresponding CMB fields are signal dominated, or when the noise is close to uniform. However, since it does not account for masking, even in these cases it is expected to be suboptimal. In practice the isotropic filter is applied to maps multiplied by an apodized mask, following Ref.~\cite{BenoitLevy:2013bc}. This maintains quasi-locality of the lensing in real space, so masking artefacts are then only expected to be significant near the mask boundaries.
For this filtering method, we mask all pixels with (temperature map) noise over $\sim9.6~\mu$K arcmin for SO and $\sim 1.8~\mu$K arcmin for S4. Before applying this new mask, it is apodized using a 30 arcmin-width Gaussian to avoid power spectrum errors due to a sharp cut-off in the sky maps. Both of these procedures result in an effective scanned area which is $\sim46\%$ smaller than the original scanned area, shown by the green outline in Fig.~\ref{fig:weights}.
The assumed isotropic noise level in the filter is taken to be that which minimizes the variance of the reconstructed lensing potential (we consider all reconstructed multipoles for this minimization as we saw no significant differences using different multipole ranges).

%%%%%%%%%%%%%%%%%%%%%%%%%%%%%%%%%%%%%%%%%%%%
\subsubsection{Isotropic filtering of weighted maps}
\label{sec:isotropic_filtering_on_weighted_maps}

Instead of applying isotropic filtering to apodized masked maps, one can also apply it to a more generally weighted map, for example to try to down-weight regions with higher noise. This corresponds to using the filter $\left(\mb C^{\fid} \mb^\top + \N\right)^{-1} \mW \vex$, where $\mW$ is diagonal in pixel space and $\N$ is taken to be isotropic. We consider the specific case where $W(\vecx,\vecx)$ is proportional to the inverse of the instrumental pixel noise variance, similar to the weighting applied by ACTpol~\cite{Sherwin:2016tyf}, so the diagonal of $\mW$ is the normalized hit count map, having values between zero (for masked areas) and one. This is expected to be close to optimal when the CMB fields are dominated by instrumental noise, so that the $\sim \N^{-1}$ dependence of the optimal filter is mostly captured by the weights.

In the case of polarization, specific window functions can also be used for separation of $E$ and $B$ modes on the partial sky~\cite{Smith:2005gi}. Several authors have
applied lensing reconstruction $EB$ estimators to the separated modes~\cite{Pearson:2014qna,Sherwin:2016tyf,Array:2016afx}. This has the advantage of being relatively fast to implement, but is clearly sub-optimal and previous analyses have not attempted to also accurately account for spatial variation in the noise. The optimal filter described in \ref{sec:optimal_filtering} already accounts for the cut-sky-induced $B$ modes, since the optimal filter automatically down-weights $B$ modes that are substantially contaminated by variance of the $E$ modes.

%%%%%%%%%%%%%%%%%%%%%%%%%%%%%%%%%%%%%%%%%%%%
\subsection{Quadratic estimators}
\label{subsec:quadratic_estimators}

We calculate the unnormalized QE $\hat{\g}_0^{\phi}(\vecx)$ using different field combinations: temperature-only (T), polarization-only (P), or temperature+polarization (minimum variance, MV). We cut lensing modes outside the range $40 \leq L\leq 3000$ in harmonic space where our reconstruction is considered unreliable.

Unless we use isotropic filtering (or even in this case close to the mask boundaries), the estimator normalization in Eq.~\eqref{eq:quad_est} is non-diagonal and difficult to calculate exactly, effectively varying spatially over the sky. We do not attempt to correctly normalize the lensing reconstruction map here, but instead correct the normalization at the level of the power spectrum (see below). If the filter varies smoothly over the map, an approximately-normalized reconstruction map could be made by using locally-defined values of the analytic isotropic normalization (see Sec.~\ref{sec:analytic} below).

%%%%%%%%%%%%%%%%%%%%%%%%%%%%%%%%%%%%%%%%%%%%
\subsection{Mean field and normalization}
\label{subsec:mean_field_and_normalization}

We calculate the mean field (MF) of the unnormalized QE ${\langle\hat{\g}^{\phi}_{0}\rangle}_{\MC}$ twice, each calculation using 50 different MC simulations. This gives two MF estimates, $\MF_1$ and $\MF_2$, with independent MC noise, so the lensing power spectrum calculated from a pair of MF-subtracted QEs will not have any MC noise bias.
Following the MF subtraction and response normalization we obtain the estimator for the lensing potential $\hat{\phi}$ given by
\beqa
\hat{\phi}_{\vecL} &\equiv& \frac{1}{\R_L^{\phi,\fid}}\left(\hat{\g}^{\phi}_{0, \vecL}-{\langle\hat{\g}^{\phi}_{0,\vecL}\rangle}_{\MC}\right),
\label{eq:phi_estimator}
\enqa
where $\R_L^{\phi,\fid}$ is a fiducial isotropic response for the reconstruction (see Eq.~\eqref{eq:analytic_response}), and $\vecL$ is the 2D flat-sky multipole vector. The definition of the isotropic response is discussed later in Sect.~\ref{sec:analytic}, along with how the power spectrum can be corrected for the anisotropy.

%%%%%%%%%%%%%%%%%%%%%%%%%%%%%%%%%%%%%%%%%%%%
\subsection{Filtering the quadratic estimators}
\label{subsec:filtering_the_quadratic_estimators}

The likelihood-based lensing power spectrum estimator given by Ref.~\cite{Hirata:2003ka} involves the CMB maps via the maximum a posteriori (MAP) estimate of the lensing field. Since MAP estimators give an estimate of the Wiener-filtered (i.e. $\mC^{\kappa\kappa}\left(\mC^{\kappa\kappa}+\mN^\kappa_{0,\rm MAP}\right)^{-1}$-filtered) field, this suggests that a close-to-optimal analysis should use estimated lensing maps weighted by their inverse covariance. More generally, if we approximate the lensing reconstruction as a Gaussian lensing field plus Gaussian reconstruction noise, the optimal power spectrum would also involve the inverse-covariance weighted fields.

When using isotropic filtering the reconstruction noise is considered to be isotropic, so an additional isotropic filtering step would simply be a re-definition of the diagonal normalization.
With optimal anisotropic filtering things are more complicated, but we can use the same iterative filtering method presented in Sec.~\ref{sec:optimal_filtering} if we approximate the lensing reconstruction noise as diagonal in pixel space. This is clearly not a good approximation in general since the lensing reconstruction noise spectrum is not white; however, if the noise is slowly varying, the large-scale effect of slowly varying reconstruction noise can be approximately modelled using a patch-uncorrelated estimate of the local noise variance $N_{0,\eff}^{\kappa}(\vecx)$. Here we use the lensing convergence ($\kappa$) reconstruction noise, since it is the convergence reconstruction which is approximately local in real space, and hence is uncorrelated between patches on large scales and has approximately white noise.
We take the effective reconstruction noise $N_{0,\eff}^{\kappa}(\vecx^{p})$ in each patch $p$ to be the average value of $N_{0,L}^{\kappa}$ (the isotropic reconstruction noise $N_{0,L}$ for $\kappa_{\vecL} \equiv L(L+1) \phi_{\vecL} / 2$ and the local patch noise value\footnote{We use $L(L+1)$ albeit the flat-sky analysis, as this factor is arbitrary when used consistently throughout.}) over the multipole range $40\leq L\leq 90$. Each patch is composed of pixels of approximately the same instrumental noise (each patch could be taken to be a single pixel, but it is more numerically convenient to model batches of pixels with similar noise levels together).

An example for the inverse of the reconstruction noise map is shown in Fig.~\ref{fig:inv_Neff_map} for lensing reconstruction from temperature and polarization. Averaging over a different multipole range to define $N_{0,\eff}^{\kappa}(\vecx)$ would result in a better agreement with the full $N_{0,L}$-filtered theory variances across corresponding multipoles, although we demonstrate that our chosen range is already quite optimal for the multipoles in which the improvement from this filtering stage is considered significant (see Fig.~\ref{fig:theoretical_variances_comparison} for comparison between using $N_{0,\eff}^{\kappa}$ and $N_{0,L}$ in the filter).

The $\mC^{\kappa\kappa}_{\fid}\left(\mC^{\kappa\kappa}_{\fid}+\mN^\kappa_{0,\eff}\right)^{-1}$ filtering process should be applied to the correctly normalized reconstructed convergence map, but the QE constructed from optimally-filtered CMB maps is not correctly normalized locally.
However, in the patch approximation, we can approximately normalize the map locally using a local response:
we take the un-normalized full QE field $\hat \g^{\kappa} (\vecx)$, and normalize locally in real space by an effective local response map $\R_{\eff}^{\kappa}(\vecx)$. This gives a reconstruction which is approximately locally normalized following Eq.~\eqref{eq:iso_quad_est}, where we define the approximate effective $\R_{\eff}^{\kappa}$ analogously to $N^\kappa_{0,\eff}$.
The final approximately filtered estimator is then
\begin{equation}
\hat{\boldsymbol{\kappa}}^{\filt} \equiv \mC^{\kappa\kappa}_{\fid}\left(\mC^{\kappa\kappa}_{\fid}+\mN^\kappa_{0,\eff}\right)^{-1} \left(\boldsymbol{\R}_{\eff}^{\kappa} \right)^{-1} \hat \vg^{\kappa},
\label{eq:kappa_filt}
\end{equation}
where\footnote{Notice that the $\phi_{\vecL}\rightarrow \kappa_{\vecL}$ conversion factor $L(L+1)/2$ is being divided instead of multiplied, as the conversion here is between the unnormalized QEs of $\phi$ and $\kappa$, so the scaling also accounts for the conversion of the response.}
\begin{equation}
\hat{\g}_{\vecL}^{\kappa} \equiv \left(\hat{\g}_{\vecL}^{\phi}-{\langle\hat{\g}_{\vecL}^{\phi}\rangle}_{\MC}\right) \times \frac{2}{L(L+1)}.
\end{equation}
Here $\mC^{\kappa\kappa}$ is diagonal in harmonic space and we have defined $\boldsymbol{\R}_{\eff}^\kappa$ and $\mN_{0,\eff}$ to be diagonal in pixel space (with $N_{0,\eff}^{\kappa}(\vecx)^{-1}=0$ in masked areas). The only non-trivial filtering step involving $\left(\mC^{\kappa\kappa}_{\fid}+\mN^\kappa_{0,\eff}\right)^{-1} $ is therefore analogous to the optimal CMB map filtering discussed in Sec.~\ref{sec:optimal_filtering}, and can be performed using the same conjugate gradient techniques.

Although the filtering is very approximate, it does not introduce any biases in the power spectrum as long as the approximate filter is propagated self-consistently into the normalization (see Sec.~\ref{sec:analytic} for more details). For simplicity we call this estimator the $\kappa$-filtered estimator.

%%%%%%%%%%%%%%%%%%%%%%%%%%%%%%%%%%%%%%%%%%%%
\subsection{Lensing power spectrum}
\label{subsec:lensing_power_spectrum}

For a pair of lensing map estimates $\hat{\phi}_1$ and $\hat{\phi}_2$ (from Eq.~\eqref{eq:phi_estimator} without filtering $\kappa$; numerical indices indicate that a different MF was used), we obtain the cross-spectrum
\beqa
C_L^{\hat{\phi}_1\hat{\phi}_2} &\equiv& \frac{1}{\fpatch n_L}\sum\limits_{\vecell {\text{ in $\vecL$ bin}}}\hat{\phi}_{1,\vecell}\hat{\phi}_{2,\vecell}^*,
\label{eq:unbiased_estimator}
\enqa
where $n_L$ is the somewhat irregular number of modes on the flat sky assigned to lensing multipole $L$ in our simulation maps and $\fpatch$ is a normalization defined to make the estimator approximately unbiased in a fiducial model.
We give an approximate analytic formula for $\fpatch$ in Sec.~\ref{sec:analytic} below;
in simple cases it can just be interpreted as an effective fractional area of our flat-sky simulation map (see Fig.~\ref{fig:weights}).

Analogous definitions apply for the $\kappa$-filtered estimator of Eq.~\eqref{eq:kappa_filt}. For brevity, in this section we only explicitly give results for $\phi$; when using the filtered $\kappa$ estimators, $\hat{\kappa}^{\filt}$, the resulting biases are related by the usual scaling.
We subtract the connected Gaussian noise bias from the power spectrum estimator using the estimator~\cite{Story:2014hni}
\begin{equation}
\begin{split}
\MCN_{0,L}^{\hat\phi\hat\phi}&= \nonumber\\
\Bigl\langle &C_{L}^{\hat{\phi}\hat{\phi}}\left[\bar{\vex}_{\MC_{1}^{\phi_1}},\bar{\vex}_{\MC_{2}^{\phi_2}},\bar{\vex}_{\MC_{2}^{\phi_2}},\bar{\vex}_{\MC_{1}^{\phi_1}}\right] \nonumber\\
+&C_{L}^{\hat{\phi}\hat{\phi}}\left[\bar{\vex}_{\MC_{1}^{\phi_1}},\bar{\vex}_{\MC_{2}^{\phi_2}},\bar{\vex}_{\MC_{1}^{\phi_1}},\bar{\vex}_{\MC_{2}^{\phi_2}}\right]\Bigr\rangle_{\MC_{1},\MC_{2}}. \nonumber\\
\end{split}
\end{equation}
Likewise we subtract the signal-dependent $\MCN_{1,L}^{\hat\phi\hat\phi}$ bias estimated using~\cite{Kesden:2003cc,Story:2014hni}
\begin{equation}
\begin{split}
\MCN_{1,L}^{\hat\phi\hat\phi} &= \\
\Bigl\langle &C_{L}^{\hat{\phi}\hat{\phi}}\left[\bar{\vex}_{\MC_{1}^{\phi_1}},\bar{\vex}_{\MC_{2}^{\phi_1}},\bar{\vex}_{\MC_{1}^{\phi_1}},\bar{\vex}_{\MC_{2}^{\phi_1}}\right] \\
+&C_{L}^{\hat{\phi}\hat{\phi}}\left[\bar{\vex}_{\MC_{1}^{\phi_1}},\bar{\vex}_{\MC_{2}^{\phi_1}},\bar{\vex}_{\MC_{2}^{\phi_1}},\bar{\vex}_{\MC_{1}^{\phi_1}}\right] \\
-&C_{L}^{\hat{\phi}\hat{\phi}}\left[\bar{\vex}_{\MC_{1}^{\phi_1}},\bar{\vex}_{\MC_{2}^{\phi_2}},\bar{\vex}_{\MC_{2}^{\phi_2}},\bar{\vex}_{\MC_{1}^{\phi_1}}\right] \\
-&C_{L}^{\hat{\phi}\hat{\phi}}\left[\bar{\vex}_{\MC_{1}^{\phi_1}},\bar{\vex}_{\MC_{2}^{\phi_2}},\bar{\vex}_{\MC_{1}^{\phi_1}},\bar{\vex}_{\MC_{2}^{\phi_2}}\right]\Bigr\rangle_{\MC_{1},\MC_{2}},
\end{split}
\end{equation}
where we use noise-free (beamed) CMB simulations, and in the first two terms each pair of MC simulations in the average have the same lensing field $\phi_1$, while the last two terms constitute a (negative) instrumental noise-free $\MCN_{0,L}$. When analysing data maps (instead of simulations), using the realization-dependent (RD) $N_{0,L}$ will account for fluctuations in realization power and correct for small errors in the assumed fiducial spectrum~\cite{Hanson:2010rp,Story:2014hni,Ade:2015zua}. We do not use $\RDN_{0,L}$ here since it is numerically expensive to calculate for 500 ``data'' simulations, but we do not expect that this would change our main results, which are based on comparisons between variances of different reconstruction methods.

The resulting power spectrum estimate is then
\beqa
\hat{C}_L^{\phi\phi} &\equiv& C_L^{\hat{\phi}_1\hat{\phi}_2} - \MCN_{0,L}^{\hat\phi\hat\phi} - \MCN_{1,L}^{\hat{\phi}\hat{\phi}}.
\label{eq:lensing_power}
\enqa

\begin{figure}%[!h]
\centering
\includegraphics[width=\columnwidth]{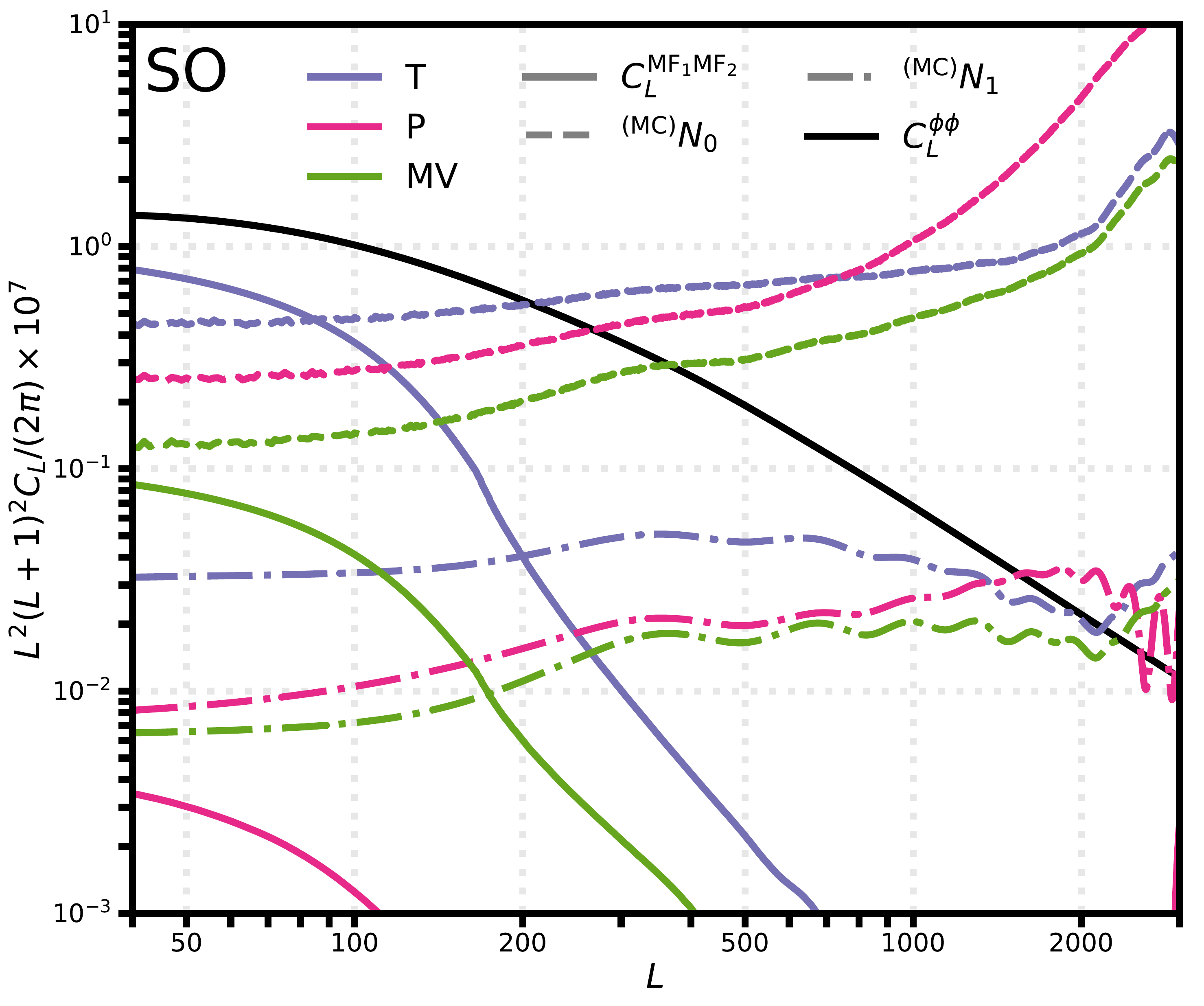}
\includegraphics[width=\columnwidth]{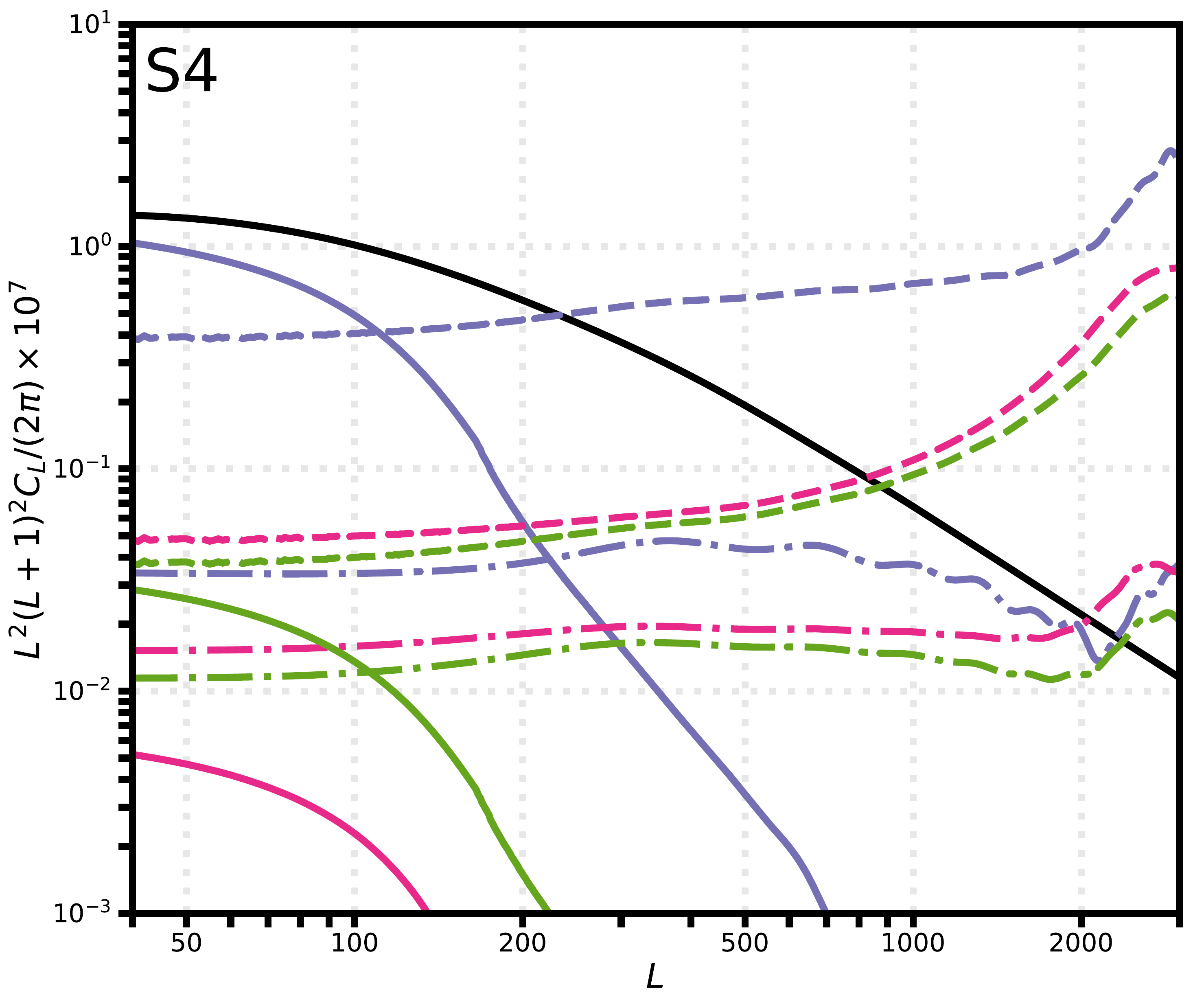}
\vspace*{-5mm}
\caption{Mean field power spectrum $C_L^{\MF_1 \MF_2}$ (solid lines), reconstruction noise $\MCN_{0,L}$ (dashed lines) and $\MCN_{1,L}$ bias (dot-dashed lines) for anisotropic-filtering analyses of temperature (green), polarization (pink) and temperature+polarization (purple) for SO (top) and S4 (bottom). The theory $C_L^{\phi\phi}$ curve (black) is shown to indicate which biases dominate the reconstruction and over which $L$ range. The MF power is the cross-correlation of the two MF values (each was obtained using 50 different simulations). The MF curves were smoothed for aesthetics, while the smoothed $\MCN_{1,L}$ shown is used in the pipeline for debiasing the reconstructed spectrum. The $N_{0,L}$ and $N_{1,L}$ curves shown for S4 are analytic using the effective noise level shown in Table~\ref{table:experiments_specs}.
}
\label{fig:biases}
\end{figure}

The de-biasing components are shown in Fig.~\ref{fig:biases} for SO- and S4-like experiments using $(\ell_{\min},\ell_{\max})=(40,3000)$ of the filtered CMB maps. We see that the temperature noise $N_{0,L}$ is similar for both SO and S4 (though this may change if a larger $\ell$-range is considered for S4). However, the S4 lensing reconstruction benefits much more from using polarization where the reconstruction is signal-dominated for a larger $L$-range. We also see the importance of MF subtraction on large scales in the presence of anisotropic noise, especially for lensing reconstructions using temperature.

%%%%%%%%%%%%%%%%%%%%%%%%%%%%%%%%%%%%%%%%%%%%
%%%%%%%%%%%%%%%%%%%%%%%%%%%%%%%%%%%%%%%%%%%%
\section{Approximate analytic model}
\label{sec:analytic}

In the case of full sky and isotropic noise, the QE of Eq.~\eqref{eq:quad_est} simplifies considerably and we can obtain analytic results. For harmonic-space field combinations
\begin{equation}
\vex \in \left\{\begin{array}{cc} T & {\rm T}\\ (E,B) & {\rm P}\\ (T,E,B) & {\rm MV}\end{array}\right.
\end{equation}
the estimator can be written as~\cite{Hu:2001kj}
\beqa
\hat{\phi}(\vecL) \equiv \left(\R^\phi_L\right)^{-1} \int\frac{d^2\vecell}{(2\pi)^2}{\vex}(\vecell-\vecL)^\dag\F_{X}(\vecell,\vecell-\vecL)\vex(\vecell),
\nonumber\\
\label{eq:iso_quad_est}
\enqa
where $\F_{X}(\vecell,\vecell-\vecL)$ collects the optimized filter and QE weights~\cite{Hu:2001kj}. The normalization (response) $\R^\phi_L$ is diagonal in harmonic space, and to obtain an unbiased estimator for cross-correlation with the true field the normalization is given by
\beqa\label{eq:analytic_response}
\R_L^\phi &=& \int\frac{d^2\vecell}{(2\pi)^2}\Tr\left[\F_{X}(\vecell,\vecell-\vecL)\vf_{X}(\vecell,\vecell-\vecL) \right],
\enqa
where the mode response functions $\vf_{X}$ are defined by
\begin{multline}
\left\la
\frac{\delta}{\delta\phi(\vecL)}\left(\vex(\vecell_1)\vex(\vecell_2)^\dag \right)
\right\ra \\= \delta(\vecell_1-\vecell_2-\vecL) \vf_X(\vecell_1,-\vecell_2).
\label{eq:moderesponse}
\end{multline}
We use gradient spectra to calculate the components of the mode response functions $\vf_{X}(\vecell_1,\vecell_2)$ following Refs.~\cite{Lewis:2011fk} and~\cite{Fabbian:2019tik}, and all theory power spectra were obtained from~\CAMB\footnote{\href{https://camb.info}{https://camb.info}}~\cite{Lewis:1999bs}. Note that the response is related to the reconstruction noise by $\R_L \sim N_{0,L}^{-1}$~\cite{Hu:2001kj}, but without exact equality here because of the use of gradient rather than lensed spectra in the mode response functions $\vf_{X}$.

In the presence of anisotropic noise and optimal filtering, the above results no longer hold, but
 we can still predict the lensing spectrum's normalization fairly accurately using the simple independent-patch approximation.
The lensing convergence QEs are all quasi-local, in that the lensing field estimated at $\vecx$ depends mostly on nearby pixels of the CMB fields where the noise level is similar if the noise is slowly varying.

Dividing the sky into patches $p$ with different approximately constant noise levels, we can define a local (correctly-normalized) isotropic estimator $\hat \phi^{p}_{\vecL}$ in each patch using the appropriate local noise level.
The full filtered estimators all locally provide estimators of the same lensing field, but have different local normalizations. We can therefore write the estimators approximately as
 \begin{equation}
\hat \phi_{\vecL} \simeq \sum_{p} w^{p}_L \hat \phi^{p}_{\vecL},
\end{equation}
where $w^{p}_L$ is a local normalization for patch $p$ and $\hat \phi^{p}_{\vecL}$ is taken to vanish outside patch $p$.
Hence, approximating the patches as uncorrelated, the power spectrum estimator of Eq.~\eqref{eq:unbiased_estimator} becomes
\beqa
C_L^{\hat{\phi}\hat{\phi}} &\simeq&\frac{1}{\fpatch n_L}\sum_{p}\sum\limits_{\vecell {\text{ in $\vecL$ bin}}} (w^{p}_L)^2
\hat{\phi}^{p}_{\vecell}{\hat{\phi}}^{p*}_{\vecell}
\\
&\simeq& \frac{1}{\fpatch}\sum_{p} (w^{p}_L)^2 f_{p} \hat{C}^{\phi^{p}\phi^{p}}_L,
\label{eq:Chatpatch}
\enqa
where $\hat{C}^{\phi^{p}\phi^{p}}_L$ is the normalized power spectrum estimator over patch $p$ and $f_p$ is the fraction of the map area in patch $p$.
To be correctly normalized after bias subtraction, this implies
\beqa
\fpatch \simeq \sum_{p} f_{p} (w^{p}_L)^2,
\label{eq:f_A}
\enqa
which gives our analytic patch approximation for the estimator normalization.
In the case of isotropic filtering with a simple binary mask, $\fpatch$ reduces to the fraction of the map area that is unmasked. The estimator is quadratic in the CMB fields, so when the filtering is an apodized mask or local weighting $W(\vecx)$, the local estimator normalization is just the square of the corresponding CMB map weight function, $w^{p}_L=[W(\vecx^{p})]^2$. The normalization $\fpatch$ then defines an $L$-independent effective map area fraction~\cite{BenoitLevy:2013bc}.\footnote{In our case, this effective area fraction equals to 0.13 and 0.03 when we mask and weight the maps respectively (corresponding to effective full-sky fractions $\fsky \approx 0.04$ and $0.01$); for the cross-correlation cases $\fpatch$ becomes 0.14 and 0.06 respectively.
}
 In the case of the cross-correlation power spectrum $C_L^{\hat{\phi}\phi}$, the corresponding normalization to be unbiased is instead
\beqa
\fpatch^{\rm cross} \simeq\sum_{p} f_{p} w^{p}_L.
\enqa

 For the case of optimal anisotropic filtering, each patch is effectively locally isotropically filtered using the appropriate local noise level. Since in Eq.~\eqref{eq:phi_estimator} we applied a single fiducial response $\R_L^{\phi,\fid}$ (with a different, somewhat arbitrary, fiducial noise level), the local estimate in a patch centred on $\vecx^{p}$ is biased by an $L$-dependent factor
\beqa
w^{p}_L = \frac{\R_L^{\phi,p}}{\R_L^{\phi,\fid}},
\label{eq:wphi}
\enqa
 where $\R_L^{\phi,p}$ is the true response according to the local noise levels in patch $p$~\cite{PL2018}. Since $\R_L^{\phi,p}$ is easily calculated analytically for each patch using Eq.~\eqref{eq:analytic_response}, this provides an approximate analytical normalization for the optimally-filtered estimator.

 Fully optimal $\kappa$-filtering would give
 \beqa
 w^{p}_L=\frac{C_{L,{\fid}}^{\kappa\kappa}}{C_{L,{\fid}}^{\kappa\kappa} + N^{\kappa,p}_{0,L}},
 \enqa
 so that each correctly normalized patch is appropriately Wiener-filtered.
 Our approximate $\kappa$-filtering instead first approximately locally normalizes the optimally-filtered reconstruction map using $\R_{\eff}^\kappa$, and then approximately Wiener filters it with white noise $ N^\kappa_{{\eff},L}$ giving
\beqa
w^{p}_L=\frac{C_{L,{\fid}}^{\kappa\kappa}}{C_{L,{\fid}}^{\kappa\kappa} + N^{\kappa,p}_{0,{\eff}}} \frac{\R^{\kappa,p}_L}{\R^{\kappa,p}_\eff}.
\label{eq:wkappa}
\enqa
The patch approximation is reminiscent of the patch lensing estimator of Ref.~\cite{Plaszczynski:2012ej}, where the authors combine lensing estimators on patches with different noise levels to optimize the signal. However, our patch approximation is only used for approximate theoretical modelling of the responses and to motivate the $\kappa$-filtering step using a locally defined effective lensing reconstruction noise. Our estimator is continuous on the observed sky, so we do not have to deal with complexities related to actually dividing the CMB maps into patches, and it should also handle filtering for mask and noise variation more accurately. Ref.~\cite{Namikawa:2014yca} also combine different patches, however their motivation is different, being focussed on how to combine observations from different overlapping experiments.

\begin{figure}[!h]
\centering
\includegraphics[width=\columnwidth]{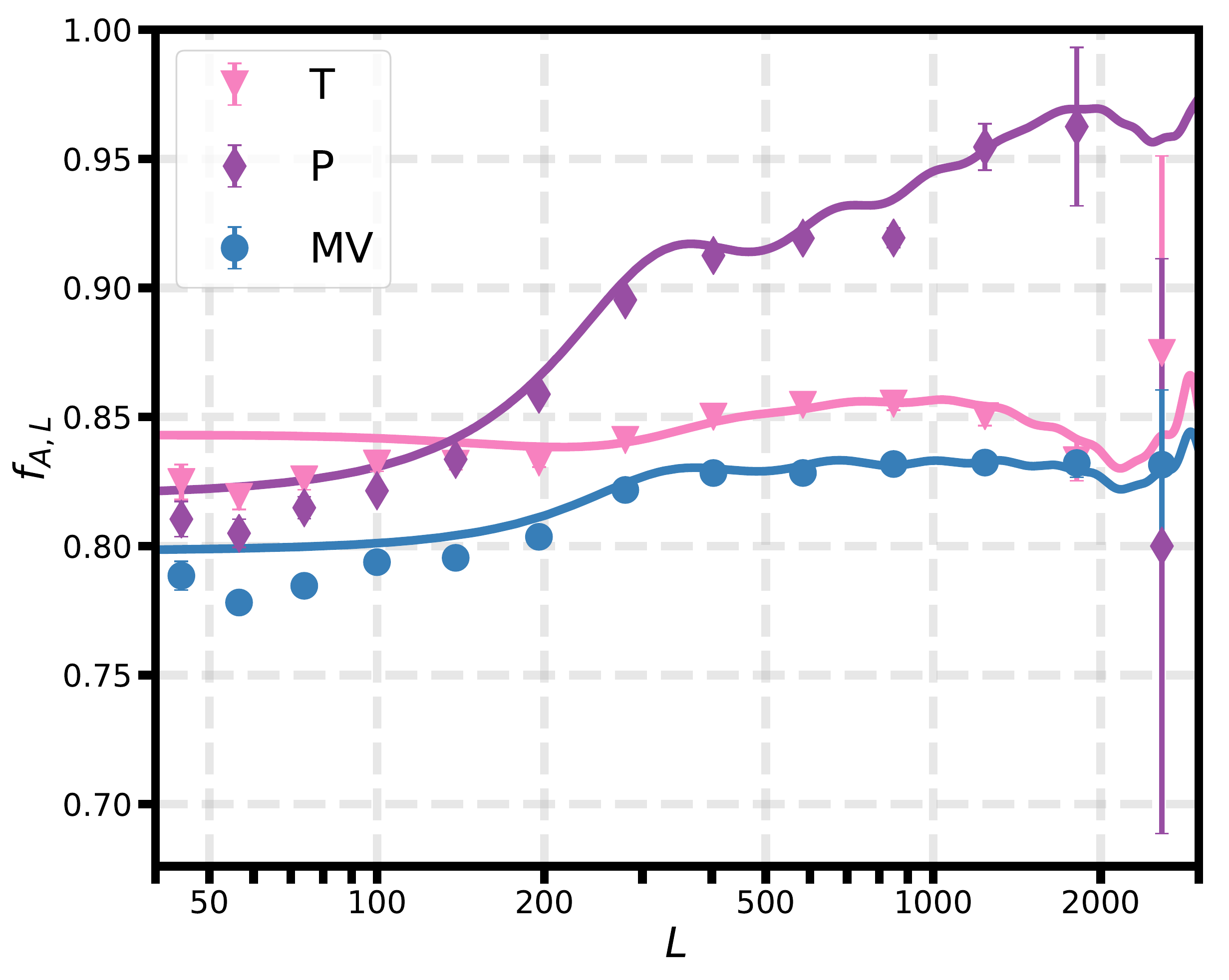}
\includegraphics[width=\columnwidth]{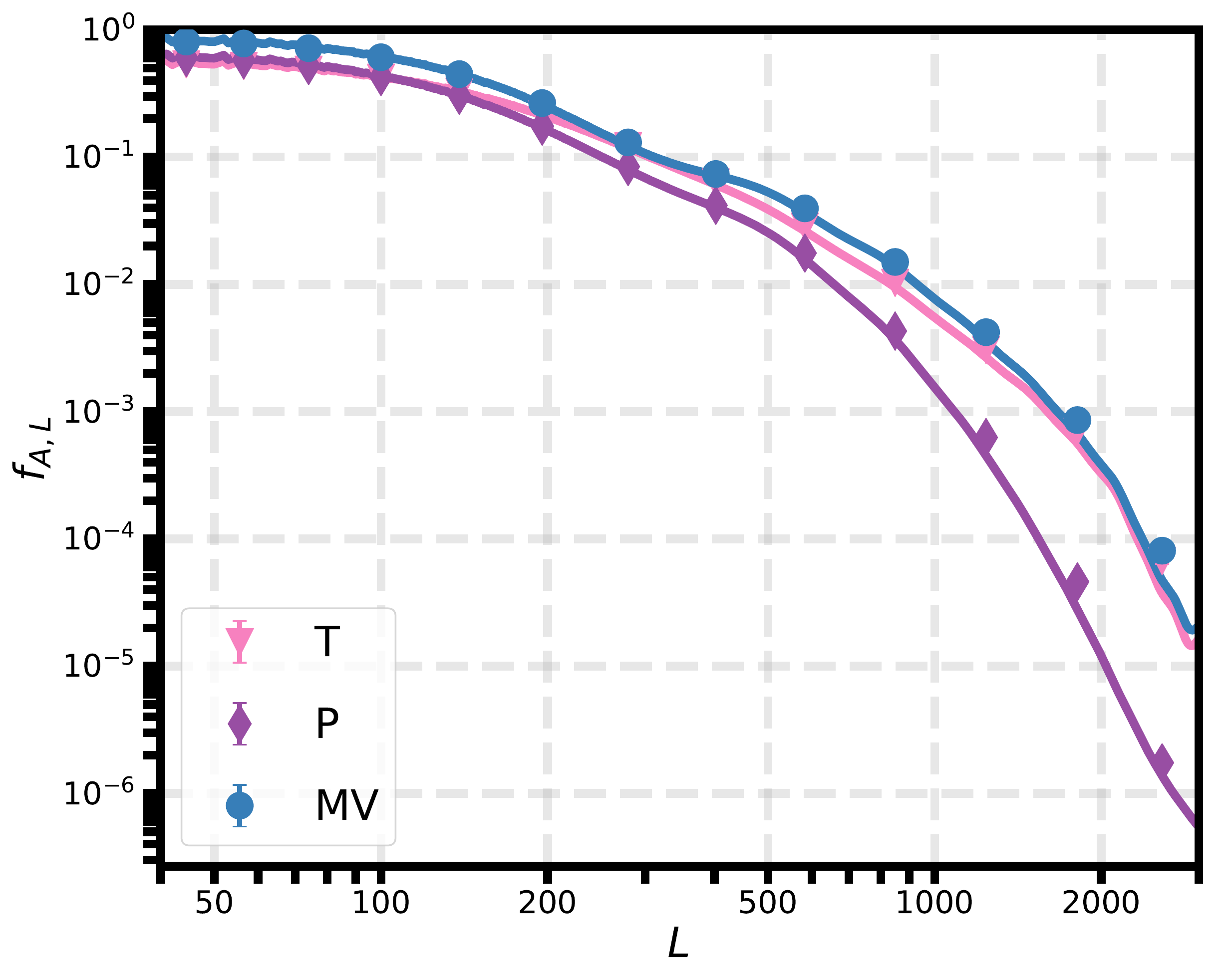}
\caption{Comparison between the approximate analytic $\fpatch$ normalization (lines) and Monte Carlo normalization (markers) required to obtain unbiased lensing reconstruction power spectra from SO-like simulations. The top plot shows the result for just optimally filtering the input CMB maps (Eq.~\eqref{eq:f_A}+\eqref{eq:wphi}), which is close to unity on all scales; its magnitude shows how close $\R_L^{\phi,\fid}$ is to the true response. The bottom plot shows the result for the estimator which is also $\kappa$-filtered (Eq.~\eqref{eq:f_A}+\eqref{eq:wkappa}), which spans many orders of magnitude due to its dependence on $C/(C+N)$ (which goes to zero on small scales but near unity on large scales). A discrepancy with the analytic result is evident on large scales for both results; the discrepancy for the $\kappa$-filtered case is shown more clearly by the additional final MC correction shown in Fig.~\ref{fig:add_MC_corr_2}.}
\label{fig:f_patch}
\end{figure}

In Fig.~\ref{fig:f_patch}, we compare the analytic correction $\fpatch$ (obtained using 160 patches in our baseline analysis, but we find the same results if we use 10 times fewer patches) with the required correction as determined from 500 MC simulations. The analytic estimate agrees well with the MC result on most scales with a slight deviation at low-$L$ (where the patch approximation is expected to break down because the real-space lensing mode size becomes comparable to the scale of variation of the noise). Similar-size MC corrections have been seen in previous analyses~\cite{PL2018}. Improving the tolerance level for convergence of the conjugate-gradient filtering does not improve this consistency: at low-$L$ the fractional difference of the $\phi$ power spectra between a tolerance of $10^{-5}$ and $10^{-6}$ for the temperature map is $\lesssim 0.1\%$. The number of patches we used to obtain the analytic correction is converged, with more patches changing the result by $\lesssim 0.1\%$.

\begin{figure}[!h]
\centering
\includegraphics[width=\columnwidth]{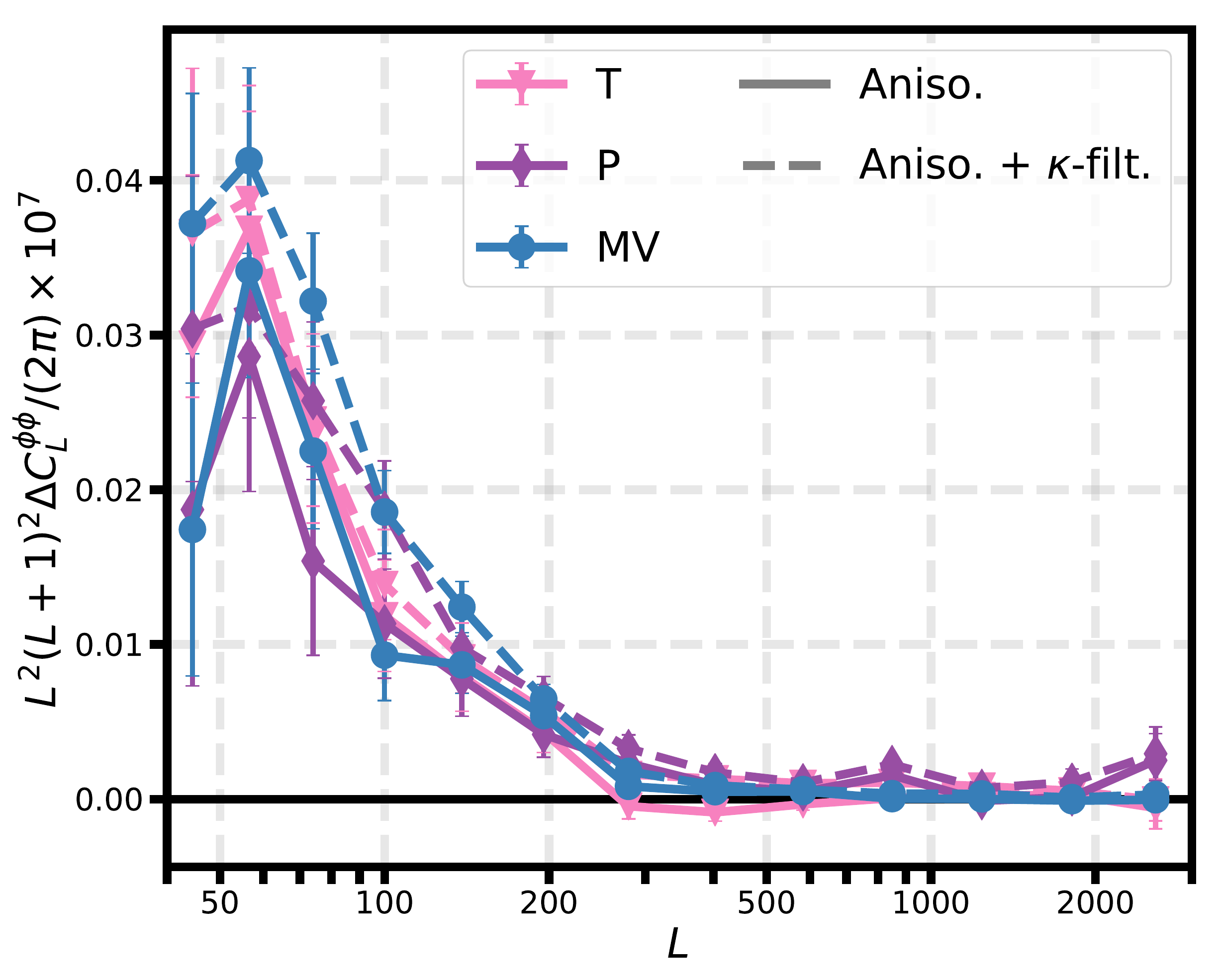}
\includegraphics[width=\columnwidth]{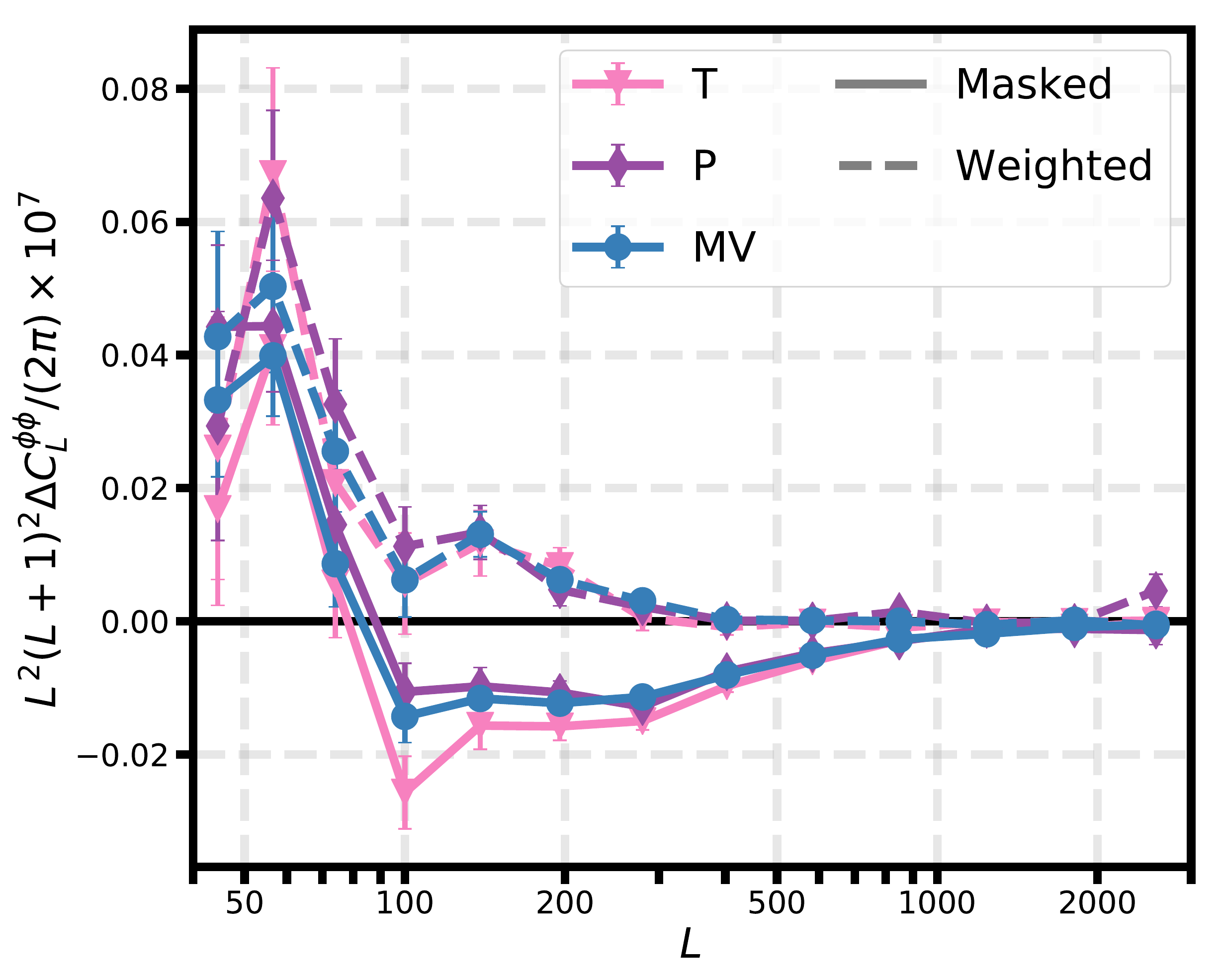}
\vspace*{-5mm}
\caption{
The additive MC correction required after applying analytic $\fpatch$ normalization for reconstruction using anisotropic filtering and combined with $\kappa$-filtering (top) and using the two isotropic filtering methods (bottom) from SO-like simulations. The MC correction for all methods is relatively small. We use this additional MC correction when plotting the final reconstruction variances in Fig.~\ref{fig:reconstruction_variances_comparison}.}
\label{fig:add_MC_corr_2}
\end{figure}

Fig.~\ref{fig:f_patch} shows that the $\fpatch$ estimates are in good agreement with simulations for $100 \lesssim L \lesssim 1500$.
To correct for the remaining inconsistency of the reconstructed power compared to the theory spectrum, especially at low-$L$, an additional small MC correction can be applied to obtain the unbiased lensing power. Fig.~\ref{fig:add_MC_corr_2} shows that the additive MC correction is close to zero over the above multipole range, and is a small but important correction elsewhere. This correction is obtained by adding $C_{L,\fid}^{\phi\phi}-\left\langle\hat{C}_L^{\phi\phi}\right\rangle$ to the reconstructed power as a final debiasing step. Since the MC correction is small, it is likely to be a good approximation to neglect its dependence on the fiducial theoretical model.

The patch approximation could also be used to obtain analytic predictions for $N_{0,L}$ and $N_{1,L}$ and then use them to debias the lensing power instead of using the MC versions. It was shown in Ref.~\cite{PL2018} that the analytic patch $N_{1,L}$ is in good agreement with $\MCN_{1,L}$ in the case of Planck. However, an MC result is required to make a reliable assessment of accuracy. In the case of $N_{0,L}$, an accurate result is critical to obtain an unbiased power spectrum: if the approximation were wrong by a few percent that would translate to a large power spectrum bias on small scales, so an accurate MC result should be used rather than the approximation. On real data, the realization-dependent $\RDN_{0,L}$ would also reduce the variance and correct leading-order sensitivity to inaccuracies in the simulations~\cite{Hanson:2010rp}. The analytic approximations should however be accurate enough to account for the model dependence of the spectra and hence construct a likelihood by straightforward generalization of the linear correction method developed by Ref.~\cite{Ade:2015zua}.

We can also use the patch approximation to assess the variance of the different estimators.
The variance of the unbiased estimator $\hat{C}^{\phi^{p}\phi^{p}}_L$ measured only over patch $p$ is given approximately by the Gaussian result
\beqa
\Var{\hat{C}^{\phi^{p}\phi^{p}}_L} \approx \frac{2(C_{L}^{\phi\phi} + N_{0,L}^{p})^2}{f_{p} n_L},
\enqa
where we neglect contributions to the variance from $N_{1,L}$ variance (which could become important on small scales) and a $p$ index indicates the value in a patch with corresponding local noise value.
Using Eq.~\eqref{eq:Chatpatch} and again taking the patches to be uncorrelated we then have
\begin{equation}
\Var{\hat C_L^{{\phi}{\phi}}} \simeq \frac{1}{[\fpatch]^2}\sum_{p} \frac{2(C_{L}^{\phi\phi} + N_{0,L}^{p})^2}{n_L} f_{p} (w^{p}_L)^4,
\end{equation}
which is approximately reduced by a factor $\Delta L$ for bins of width $\Delta L$ centred at $L$.

Without $\kappa$-filtering the expected (binned) power variance from optimally filtering the CMB maps is then
\beqa
\Var{\hat C_L^{\kappa\kappa}} = \frac{\sum\limits_{p} \frac{2(C_L^{\kappa\kappa}+N_{0,L}^{\kappa,p})^2}{\Delta L n_L} f_{p}\left(\frac{\R_L^{\kappa,p}}{ \R_L ^{\kappa,\fid}}\right)^4}{\left[\sum\limits_{p} f_{p} \left(\frac{\R_L^{\kappa,p}}{\R_L^{\kappa,\fid}}\right)^2\right]^{2}}.
\label{eq:variance}
\enqa
The best-case expected variance after filtering $\kappa$ using the ideal full $L$-dependency of $N_{0,L}^{\kappa}$ is
\beqa
\Var{\hat C_{L,\filt}^{\kappa\kappa}} = \frac{\sum\limits_{p} \frac{2(C_L^{\kappa\kappa}+N_{0,L}^{\kappa,p})^2}{\Delta L n_L} f_{p} \left(\frac{C_{L,\fid}^{\kappa\kappa}}{C_{L,\fid}^{\kappa\kappa}+N_{0,L}^{\kappa,p}}\right)^4}{\left[\sum\limits_{p} f_{p} {\left(\frac{C_{L,\fid}^{\kappa\kappa}}{C_{L,\fid}^{\kappa\kappa}+N_{0,L}^{\kappa,p}}\right)}^2\right]^{2}}. \nonumber\\
\label{eq:variance_with_kappa_filtering}
\enqa
Instead, using the effective noise level (which is what we use in practice), the expected variance becomes (from Eq.~\eqref{eq:wkappa})
\beqa
\Var{\hat C_{L,\filt}^{\kappa\kappa}} = \frac{\sum\limits_{p} \frac{2(C_L^{\kappa\kappa}+N_{0,L}^{\kappa,p})^2}{\Delta L n_L} f_{p} \left(\frac{\R_L^{\kappa,p} }{\R_{\eff}^{\kappa,p}} \right)^4 \left(\frac{C_{L,\fid}^{\kappa\kappa}}{C_{L,\fid}^{\kappa\kappa}+N_{0,\eff}^{\kappa,p}}\right)^4}{\left[\sum\limits_{p} f_{p} \left(\frac{\R_L^{\kappa,p}}{\R_{\eff}^{\kappa,p}} \right)^2 {\left(\frac{C_{L,\fid}^{\kappa\kappa}}{C_{L,\fid}^{\kappa\kappa}+N_{0,\eff}^{\kappa,p}}\right)}^2\right]^{2}}. \nonumber\\
\label{eq:variance_with_kappa_filtering_n_eff}
\enqa

\begin{figure}[!h]
\centering
\includegraphics[width=\columnwidth]{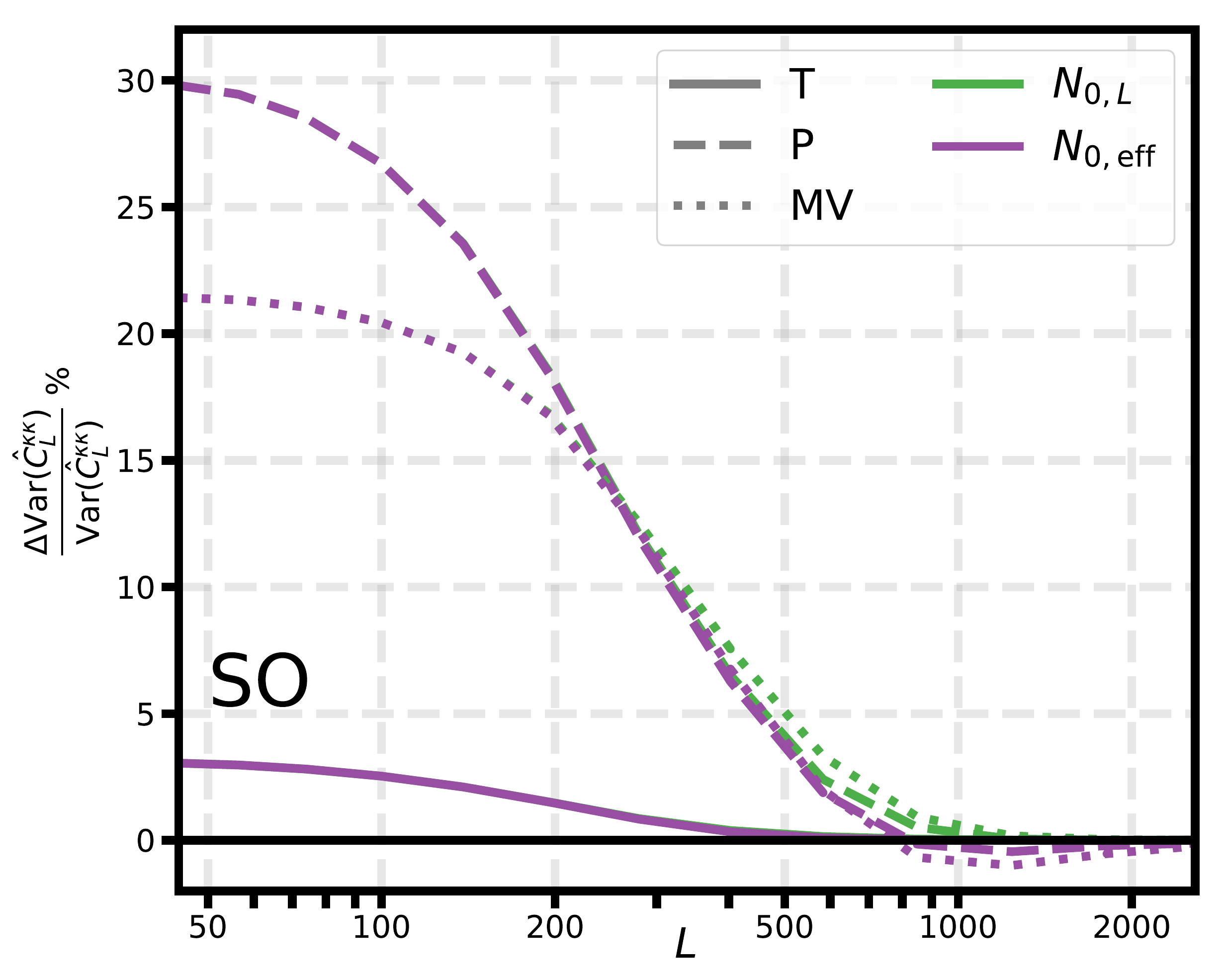}
\includegraphics[width=\columnwidth]{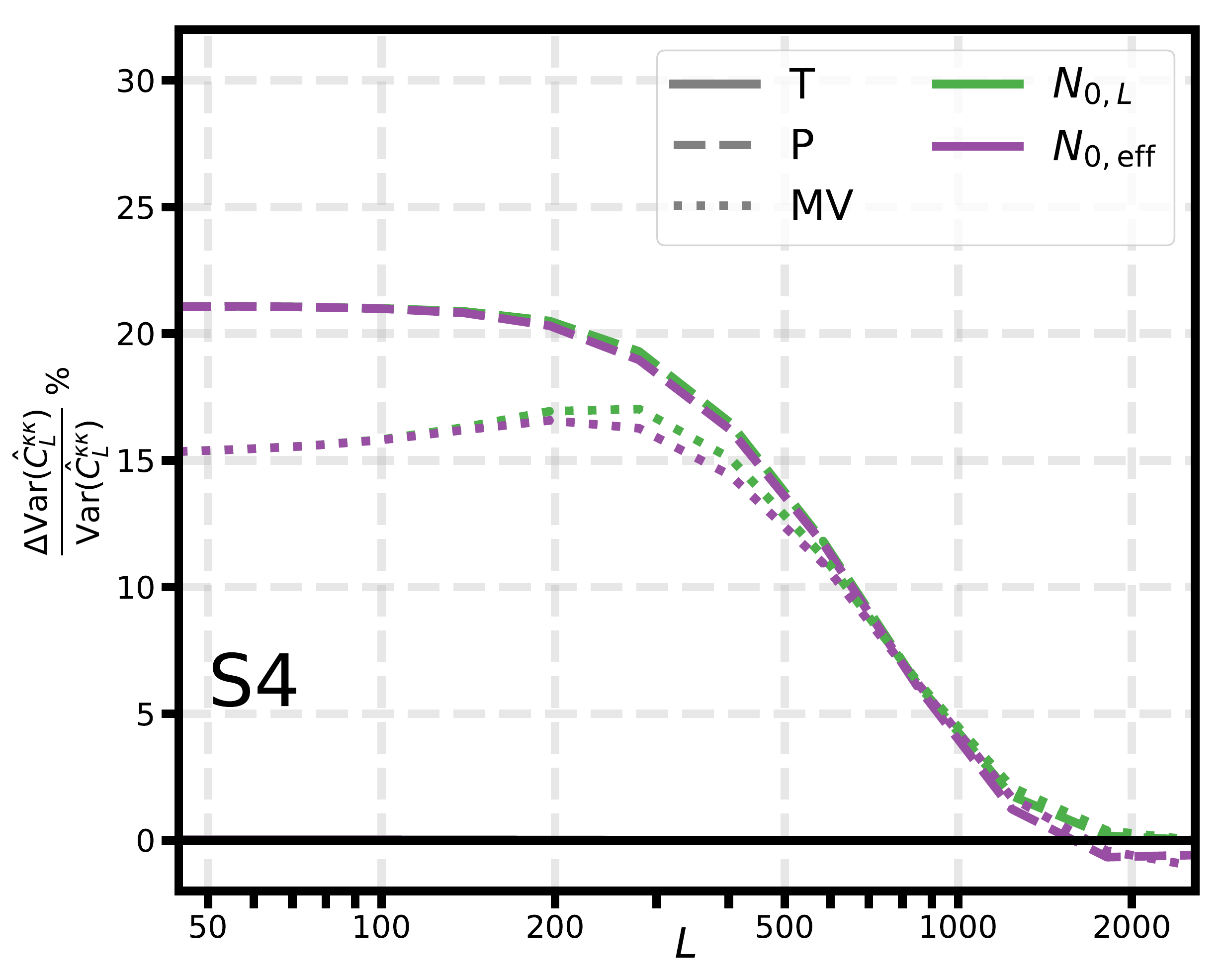}
\vspace*{-5mm}
\caption{Fractional improvement in the theoretical binned variance of $\hat{C}_L^{\kappa\kappa}$ when using an additional $\kappa$-filtering step for an SO-like experiment (top) and an S4-like experiment (bottom), estimated using an analytic patch approximation with 64 patches.
The green curves show the optimal result which would be obtained if we could locally filter with $N_{0,L}^{p}$ (Eq.~\eqref{eq:variance_with_kappa_filtering}), while the purple curves show the very similar result from our approximate local normalization followed by filtering using an effective white reconstruction noise $N_{0,\eff}^{\kappa}(\vecx)$ in the filter (Eq.~\eqref{eq:variance_with_kappa_filtering_n_eff}). The reconstruction from polarization alone benefits most from this additional filtering step since it is most affected by the noise anisotropy.}
\label{fig:theoretical_variances_comparison}
\end{figure}

The fractional differences between the unfiltered variance and the two filtered variances are shown (in percent) in Fig.~\ref{fig:theoretical_variances_comparison}. The $\kappa$-filtering step significantly reduces the variance of the large-scale power spectrum, while having little effect on small scales where just optimally-filtering the CMB maps is already nearly optimal (because the local response is effectively automatically inverse-noise weighting, and inverse-noise weighting is optimal when the noise is large).

The improvement expected from $\kappa$-filtering depends on the CMB noise level relative to the signal, and is therefore different for temperature and polarization.
For temperature analysis with our choice of $\ell_{\max}$ the improvement for SO-like noise is small, and negligible for S4, since the CMB temperature is signal dominated in both cases (so the lensing responses are already nearly isotropic because the contribution from noise variance is small). For polarization the gains are significant because the polarization noise (and hence its anisotropy) is significant in both experiments, with a somewhat larger improvement for SO where the noise is relatively more important.

The impact of both filtering steps compared to other methods will depend on the specific hit count distribution and hence relative importance of noise variations. For larger areas (such as the survey areas available for lensing in SO and S4) the counts may be more or less anisotropic than in the patch we tested in detail, leading to somewhat larger or smaller overall gains respectively.

%%%%%%%%%%%%%%%%%%%%%%%%%%%%%%%%%%%%%%%%%%%%
%%%%%%%%%%%%%%%%%%%%%%%%%%%%%%%%%%%%%%%%%%%%
\section{Results}
\label{sec:results}

We now compare simulation-based lensing reconstruction results from the different filtering methods described in Sections~\ref{subsec:filtering_methods} and~\ref{subsec:filtering_the_quadratic_estimators}.
We compare power spectrum results from applying an isotropic filter on masked or weighted CMB maps, as used by several previous experiments, to applying an optimal anisotropic filtering with or without also filtering the reconstructed $\kappa$ map. The reconstructions were made from simulations with SO- and S4-like noise and beam while for simplicity considering the same scanning strategy, corresponding to a sky fraction of $\fsky=0.13$ and strongly anisotropic hit count map shown in Fig.~\ref{fig:weights}. The instrument sensitivities, (effective) noise levels, beam widths and observation time considered for each experiment are given in Table~\ref{table:experiments_specs}. We assume an observation efficiency of $1/5$ for both experiments, and each simulation has 4096 pixels on a side with 1.7 arcminute pixel size.

\begin{table}[h!]
\begin{center}
\begin{tabular}{ccccc}
\hline
\hline
 Experiment 	& $\s_{_T}$	 			& $\Delta_T$			& $\theta_{\FWHM}$ 	& $t_{\obs}$	\\
 					& [$\mu$K-$\sqrt{\sec}$] 	& [$\mu$K-arcmin]	& [arcmin]					& [years]		\\
 \hline
  SO 				& 6.7 								& 5.0 						& 1.4 							& 5			\\
  S4 				& 1.5	 							& 1.0 						& 1.5 							& 7			\\
 \hline
 \hline
\end{tabular}
\end{center}
\caption{Experimental specifications for our SO- and S4-like simulations. The $\Delta_T$ effective map-level sensitivity is not used for the simulations, but obtained from the power spectrum of weighted noise map realizations. The value of SO's temperature sensitivity $\s_T$ is the LAT `baseline' level for 145 GHz from~\cite{Ade:2018sbj} and the corresponding $\Delta_T$ is the result from the specific hit count map considered. For S4, on the other hand, we determined the sensitivity for 145 GHz so that $\Delta_T$ results in the forecast value from~\cite{Abazajian:2016yjj,2019arXiv190801062A}\protect\footnotemark. Polarization sensitivity is taken to be $\Delta_P=\sqrt{2}\Delta_T$. We consider the same scanning time efficiency of $1/5$ and (post-filtering) CMB multiple range $(\ell_{\min},\ell_{\max})=(40,3000)$ for both experiments.}
\label{table:experiments_specs}
\end{table}

We perform lensing reconstructions from different field combinations, T, P, and MV, using 500 simulations, from which we calculate the reconstruction variances and the cross-correlation coefficients with the input lensing map.

\begin{figure}[!h]
\centering
\includegraphics[width=\columnwidth]{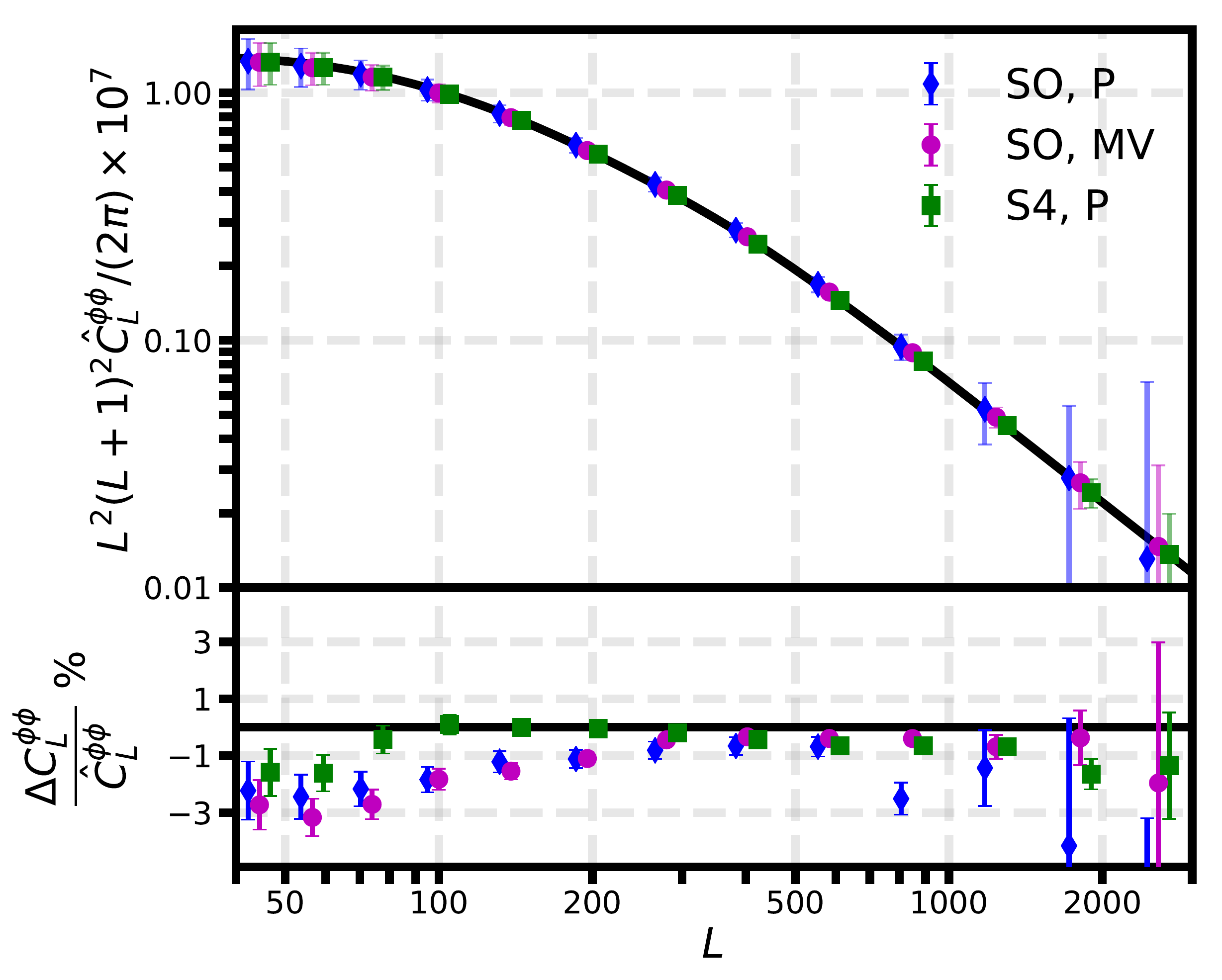}
\caption{
Simulated lensing potential reconstruction power spectrum and residuals from optimal anisotropic filtering of the CMB maps and approximate $\kappa$-filtering, for the SO MV estimator (magenta) and using only polarization maps for SO (blue) and S4 (green). An MC correction can be further applied to get the final unbiased power; see Fig.~\ref{fig:add_MC_corr_2}. The opaque error bars in the residual plot are the MC errors expected from the average of 500 simulations, while the translucent error bars in the upper panel are the latter scaled by $\sqrt{500}$ to show the uncertainty for one lensing realization (points offset for clarity, and errors slightly underestimated on small scales because we neglect the $\MCN_{1,L}$ MC error).
}
\label{fig:lensing_reconstruction_power}
\end{figure}

The reconstructed power before applying an additive MC correction is shown in Fig.~\ref{fig:lensing_reconstruction_power} from SO using P and MV, and from S4 using P. These spectra were obtained using the optimal anisotropic filter followed by filtering the reconstructed $\kappa$ map, and demonstrate that
the analytic patch normalization described in Sec.~\ref{sec:analytic} is accurate to the percent level. Comparing the two SO power spectra, we see that while the low-$L$ bias is relatively similar as expected (as this results from the mask/scan area), the high-$L$ bias is much less significant for MV, most likely due to the contribution from the signal-dominated T. To obtain the unbiased spectra, we apply small additional MC corrections. These additional corrections are shown in Fig.~\ref{fig:add_MC_corr_2}, and for both experiments are smaller than 3\%.

\begin{figure*}
\centering
\includegraphics[width=0.9\textwidth]{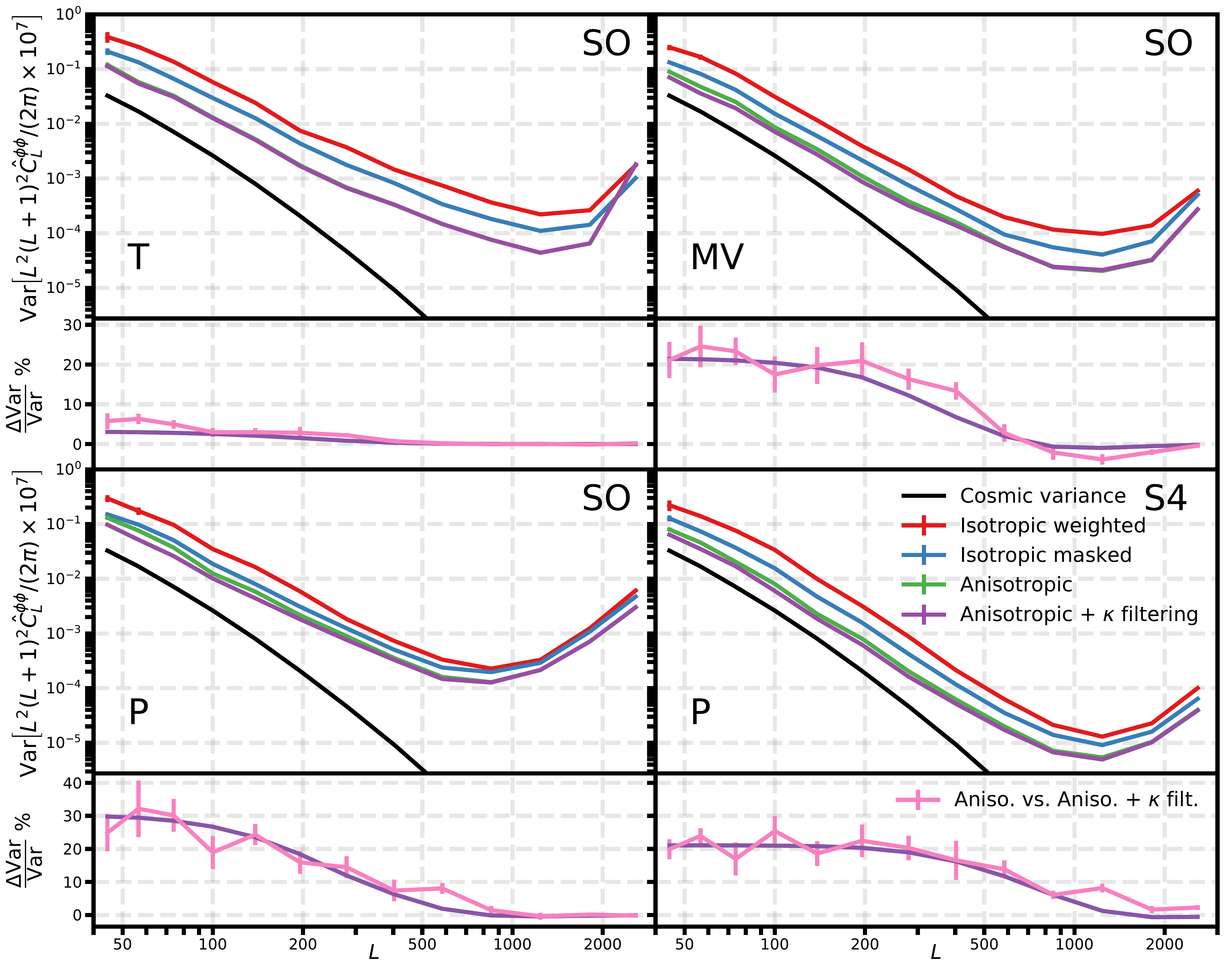}
\caption{Reconstructed lensing power spectrum variances from the various filtering methods for an SO-like experiment and an S4-like experiment (bottom right). The error bars on the variance and the fractional differences were estimated using 10 sub-batches of variance estimates. The black curve is the cosmic variance for the lensing power shown as a reference for the minimal variance we could obtain with no noise over the entire scanned area. Using the optimal anisotropic filter leads to variance improvements by a factor of 2-5 for most of the considered L-range compared to considering isotropic filtering over a reduced sky area. The improvement is largest for reconstructions using polarization, which is less signal-dominated than temperature. The lower panels of each plot show the fractional difference between reconstruction variances with and without additional $\kappa$ reconstruction filtering (pink), plotted against the corresponding theoretical curves from Fig.~\ref{fig:theoretical_variances_comparison} (purple lines), demonstrating good agreement with the approximate analytic model.
For S4, polarization dominates the reconstruction, so we only show the polarization results.
 }
\label{fig:reconstruction_variances_comparison}
\end{figure*}

The unbiased lensing reconstruction power spectra variances for the various filtering methods are shown in Fig.~\ref{fig:reconstruction_variances_comparison}. As expected, lower variances are achieved when using both temperature and polarization maps for the reconstruction. For reconstructions using temperature, the optimal anisotropic filter yields an improvement in variance by a factor of 2-5 compared to isotropic filtering a masked or weighted map, though the level of improvement does depend on the mask chosen for the isotropic-filtering analysis.
The improvement is smaller for a polarization-only reconstruction where reconstruction noise rather than cosmic variance is relatively more important near the edges of the scanned area. The isotropic filtering results depend both on the noise level used in the isotropic filter, which was chosen to minimise the variance, and (in the masked case) also the masked area actually used (to reduce variance from very noisy areas near the edge). We did not optimize the mask area, but testing with several sensible masking schemes showed no large variance improvement. Compared to the isotropic filtering methods, the optimal and $\kappa$-filtering methods have fewer free parameters, and varying these only has a small effect on results, which are already close to optimal.

The difference plots in Fig.~\ref{fig:reconstruction_variances_comparison} show the further fractional improvement in variance after applying our approximate additional $\kappa$ reconstruction filtering. Results agree well with the predicted theoretical curves shown in Fig.~\ref{fig:theoretical_variances_comparison}, demonstrating that the patch approximation is capturing the main effect well. The predicted $\sim 30\%$ reduction in variance on large scales is therefore achievable in practice, at only a small additional numerical cost.

\begin{figure}
\centering
\includegraphics[width=\columnwidth]{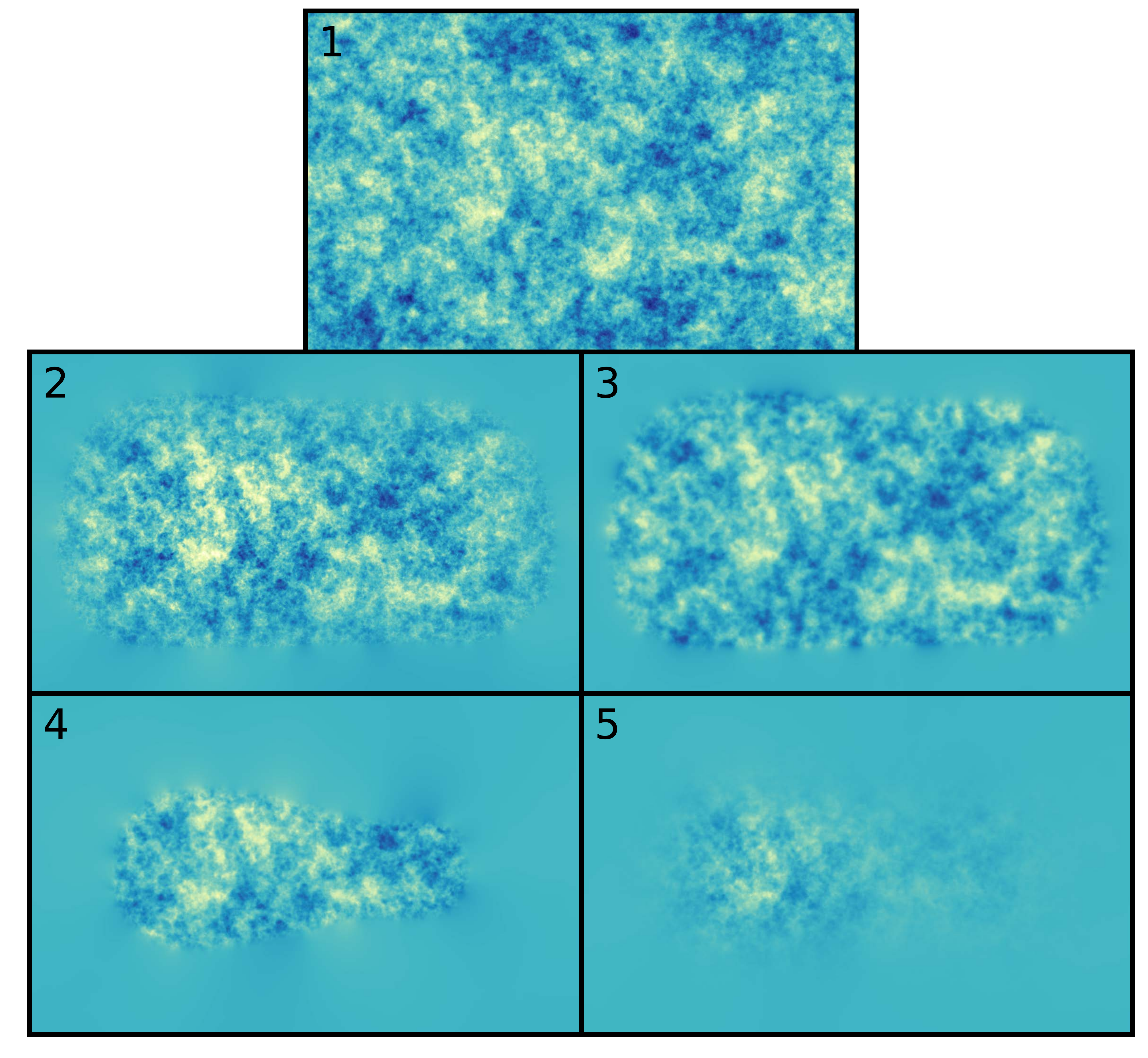}
\caption{
\one: Input deflection $\alpha$ map of one of the realizations used (where $\alpha_{\vecL}=\sqrt{L(L+1)}\phi_{\vecL}$).
\two: The Wiener-filtered reconstructed $\alpha$ obtained when applying an optimal anisotropic filter to the temperature and polarization maps. The map is normalized using a fiducial isotropic normalization $\R_L^{\fid}$, where the specific choice of effective isotropic noise chosen has some impact on the map, and the isotropic WF applies $\frac{C_{L,\fid}^{\phi\phi}}{C_{L,\fid}^{\phi\phi}+\MCN_{0,L}^{\phi}}$ to $\hat\phi_{\vecL}$.
\three: The $\alpha$ reconstruction map from applying a further anisotropic filter on the optimally-reconstructed $\kappa$ map after approximate local normalization.
The approximate white local normalization improves the match to the input near the patch boundaries, but underweights small scales compared to the true normalization ($\R^\kappa_L\sim N^\kappa_{0,L}{}^{-1}$ falls at high L compared to fixed value we chose that matches on large scales), so this map appears smoother. This is corrected at the power spectrum level by the analytic patch normalization correction.
\four: The Wiener-filtered reconstructed $\alpha$ map using an isotropic filter on a masked map. The same WF as in panel 2 is applied with the respective $\MCN_{0,L}^{\phi}$.
\five: The Wiener-filtered reconstructed $\alpha$ map using an isotropic filter on a weighted map. The same WF as in panel 2 is applied with the respective $\MCN_{0,L}^{\phi}$.
All reconstructed maps are from the same lensing potential realization, and show the same colour ranges as the input map.
}
\label{fig:lensing_reconstruction_maps}
\end{figure}

We also compare some reconstructed real-space lensing maps from the various reconstruction method. Fig.~\ref{fig:lensing_reconstruction_maps} shows the input $\alpha$ map (where $\alpha_{\vecL}=\sqrt{L(L+1)}\phi_{\vecL}$) of one simulation in comparison to the MV reconstruction maps using the same lensing realization with the various different filtering methods.
Qualitatively, all maps demonstrate a good reconstruction in the low-noise pixels near the centre of the patch that have longer observation times.
Isotropic filtering on weighted maps significantly down-weights the reconstruction in the higher-noise pixels, while the reconstruction from masked maps only down-weights around the edges of the mask (however the effective area is reduced due to the mask excluding high noise pixels). When applying the optimal anisotropic filter, we reconstruct the lensing potential on most of the scanned region giving lower power spectrum variance. Applying the further $\kappa$-filtering step on the locally normalized map removes the effective down-weighting around the edges of the scanned sky area, which can explain the variance improvements we see at low-$L$ where the reconstruction is signal-dominated.

We then calculate the cross-correlation coefficients with the input lensing map,
\begin{equation}\label{eq:cross_correlation_coefficient}
\hat \rho_{_L} \equiv \frac{C_L^{ \hat{\phi}\phi}}{\sqrt{C_L^{\hat{\phi}\hat{\phi}}C_L^{\phi\phi}}}.
\end{equation}

\begin{figure}[!h]
\centering
\includegraphics[width=\columnwidth]{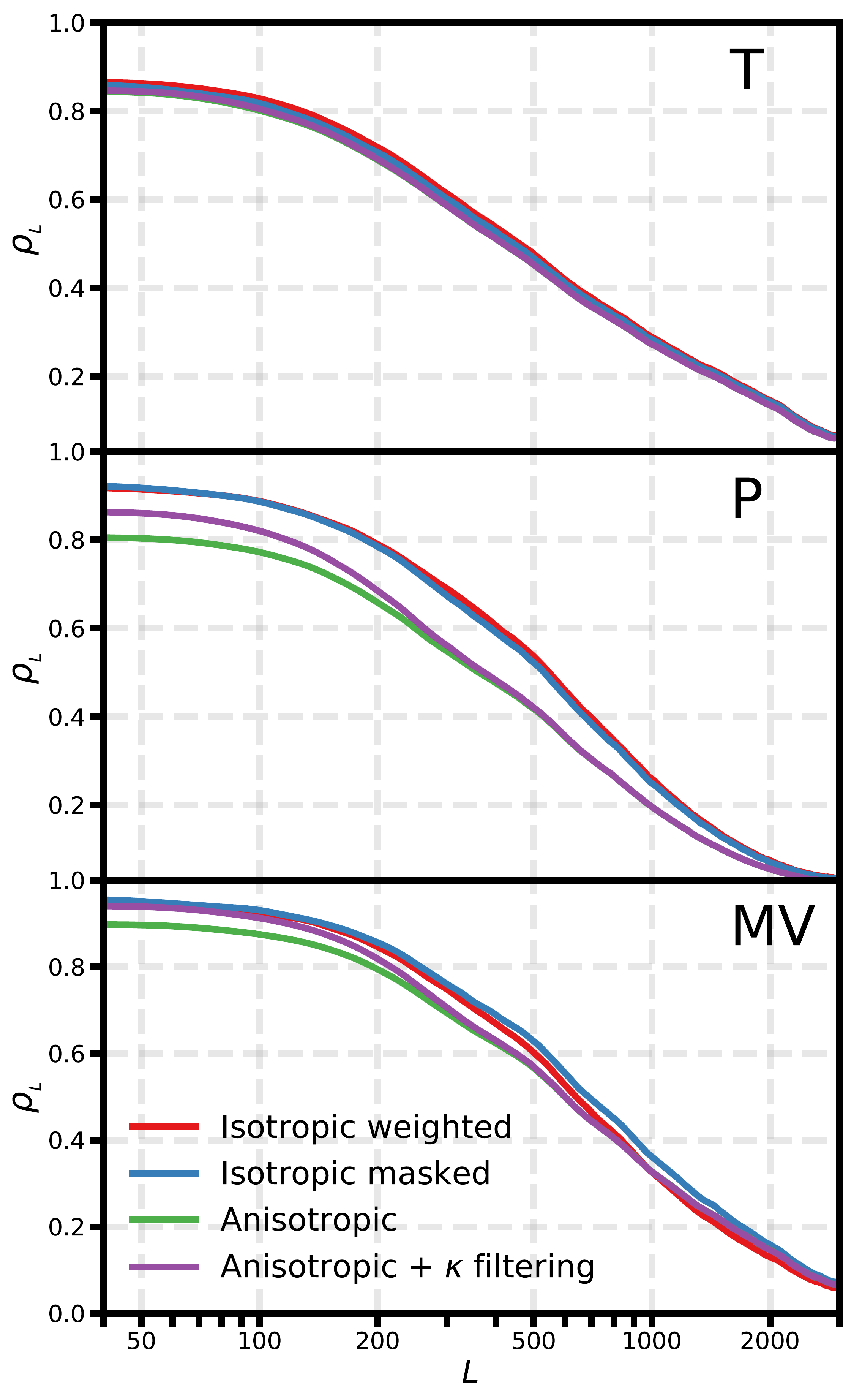}
\caption{Cross-correlation coefficients (Eq.~\eqref{eq:cross_correlation_coefficient}) for the various reconstruction methods and an SO-like experiment. The curves were smoothed with a $\sigma_L = 12$ width Gaussian to reduce sampling noise. The best reconstruction is made from a combined analysis of temperature and polarization maps (MV). The improvement due to the additional $\kappa$-filtering step when using anisotropic filter on the CMB maps is mostly visible for $L\lesssim200$. The most correlated maps are those obtained from applying an isotropic filter over a reduced sky area, but this reduced sky area loses information compared to other methods using the full observed area.
}
\label{fig:cross_correlation_coefficient}
\end{figure}

The cross-correlation coefficients of the SO reconstructions are shown in Fig.~\ref{fig:cross_correlation_coefficient}. We see an improved correlation after filtering the reconstructed $\kappa$ map compared to only anisotropically filtering the CMB maps. This improvement is most visible for the signal-dominated regime at $L\lesssim300$. The lensing reconstruction from the isotropic filtering process is done over a smaller effective area, in which the signal-to-noise is already high, so the cross-correlation coefficient restricted to that area is higher (over the same area the more optimal methods would also give substantially higher cross-correlation coefficients).

\footnotetext{Reference design is also available at: \href{https://cmb-s4.org/wiki/index.php/Expected\_Survey\_Performance\_for\_Science\_Forecasting\#DSR\_Reference\_Design}{https://cmb-s4.org/wiki/index.php/\\Expected\_Survey\_Performance\_for\_Science\_Forecasting}.}

%%%%%%%%%%%%%%%%%%%%%%%%%%%%%%%%%%%%%%%%%%%%
%%%%%%%%%%%%%%%%%%%%%%%%%%%%%%%%%%%%%%%%%%%%
\section{Conclusions}
\label{sec:conclusions}

In this paper we demonstrated the importance of optimizing the reconstruction pipeline to minimize the lensing spectrum variance when sky maps have anisotropic noise. We showed that optimal CMB map filtering can have significant gains compared to simple isotropic filtering (a factor of 2-5 decrease in variance for our choice of masking on the configurations tested).

Lensing reconstructions using optimally filtered maps are effectively inverse-noise weighted because the normalization response is directly related to the reconstruction noise in simple cases. This weighting is nearly optimal for the power spectrum on small scales where the reconstruction is noise dominated, however it is significantly suboptimal on larger scales where lensing modes are reconstructed with high signal-to-noise. We showed that an additional approximate $\kappa$-reconstruction filtering stage can significantly improve the variance of power spectrum estimates in the signal dominated regime on large scales, while also smoothly approaching close to the original optimal result on small scales. The anisotropic filtering performs well for both SO- and S4-like noise levels, and for the specific anisotropic noise we tested the $\kappa$ filtering step reduces the variance by about 30\% on large scales.
Our optimal filtering steps use a conjugate gradient approach, making the optimized estimators easily numerically tractable (but still somewhat numerically expensive; there is potential for further gains using other methods, e.g. filtering using a pre-trained neutral network~\cite{Munchmeyer:2019kng}).

We used a flat-sky analysis, but do not expect our results to be significantly different when applying to a full-sky lensing reconstruction, though developing a full-sky analysis is clearly a requirement for analysis of realistic data over large sky areas. We also considered only a single frequency map and ignored the complication of foreground residual modelling. The relative improvements that we have shown should however still remain valid as they only depend on the overall broad distribution of the map hit counts.
Realistic ground-based data also usually has strongly correlated noise, making full signal-plus-noise filtering substantially more challenging; however, the amplitude of the noise would still follow the broad hit count distribution, so the uncorrelated noise approximation that we made may still be sufficient to obtain significant gains compared to less optimized estimators. Further work would be needed to study the best way to filter correlated noise in practice.

The fast optimized QE lensing pipeline presented in this work can easily be run on many simulations, and so may prove valuable for quantifying the impact of systematics, foregrounds, and other effects that can be simulated on lensing reconstruction. The effect of several potentially dangerous systematics will be explored in an upcoming paper. The simple but accurate analytic patch approximations results may also prove valuable for optimization of observing strategies.

Although we have demonstrated significant gains compared to simple QE estimators, our results are clearly not fully optimal both because of approximations in the $\kappa$-filtering and because the estimator is still fundamentally quadratic. A likelihood based approach using iterative estimators~\cite{Hirata:2003ka,Carron:2017mqf} could perform substantially better in the high signal-to-noise regime where quadratic estimators become suboptimal. However, a fully optimal power spectrum estimator applicable to realistic cut-sky data with inhomogeneous noise does not currently exist, and developing such an estimator would be an interesting avenue for future research. Comparison with a fully optimal estimator would allow us to assess gains compared to the approximate estimators we have presented. However, our approximate estimators are likely to still remain useful as they are fast to calculate and straightforward to approximately model analytically using the patch approximation.

%%%%%%%%%%%%%%%%%%%%%%%%%%%%%%%%%%%%%%%%%%%%
%%%%%%%%%%%%%%%%%%%%%%%%%%%%%%%%%%%%%%%%%%%%
\section{Acknowledgements}
We thank Neil Goeckner-Wald for providing us with the scanning strategy hit count map for simulating realistic anisotropic noise. We acknowledge support from the European Research Council under the European Union's Seventh Framework Programme (FP/2007-2013) / ERC Grant Agreement No. [616170] and AL support by the UK STFC grant ST/P000525/1.
This research used resources of the National Energy Research Scientific Computing Center (NERSC), a U.S. Department of Energy Office of Science User Facility operated under Contract No. DE-AC02-05CH11231. Some of the results in this paper have been derived using the HEALPix~\cite{Gorski:2004by} package.

\indent

%%%%%%%%%%%%%%%%%%%%%%%%%%%%%%%%%%%%%%%%%%%%
%%%%%%%%%%%%%%%%%%%%%%%%%%%%%%%%%%%%%%%%%%%%
% References

\bibliography{lensingbib,texbase/antony}

\providecommand{\aj}{Astron. J. }\providecommand{\apj}{ApJ
  }\providecommand{\apjl}{ApJ
  }\providecommand{\mnras}{MNRAS}\providecommand{\prl}{PRL}\providecommand{\prd}{PRD}\providecommand{\jcap}{JCAP}\providecommand{\aap}{A\&A}
\begin{thebibliography}{33}%
\makeatletter
\providecommand \@ifxundefined [1]{%
 \@ifx{#1\undefined}
}%
\providecommand \@ifnum [1]{%
 \ifnum #1\expandafter \@firstoftwo
 \else \expandafter \@secondoftwo
 \fi
}%
\providecommand \@ifx [1]{%
 \ifx #1\expandafter \@firstoftwo
 \else \expandafter \@secondoftwo
 \fi
}%
\providecommand \natexlab [1]{#1}%
\providecommand \enquote  [1]{``#1''}%
\providecommand \bibnamefont  [1]{#1}%
\providecommand \bibfnamefont [1]{#1}%
\providecommand \citenamefont [1]{#1}%
\providecommand \href@noop [0]{\@secondoftwo}%
\providecommand \href [0]{\begingroup \@sanitize@url \@href}%
\providecommand \@href[1]{\@@startlink{#1}\@@href}%
\providecommand \@@href[1]{\endgroup#1\@@endlink}%
\providecommand \@sanitize@url [0]{\catcode `\\12\catcode `\$12\catcode
  `\&12\catcode `\#12\catcode `\^12\catcode `\_12\catcode `\%12\relax}%
\providecommand \@@startlink[1]{}%
\providecommand \@@endlink[0]{}%
\providecommand \url  [0]{\begingroup\@sanitize@url \@url }%
\providecommand \@url [1]{\endgroup\@href {#1}{\urlprefix }}%
\providecommand \urlprefix  [0]{URL }%
\providecommand \Eprint [0]{\href }%
\providecommand \doibase [0]{http://dx.doi.org/}%
\providecommand \selectlanguage [0]{\@gobble}%
\providecommand \bibinfo  [0]{\@secondoftwo}%
\providecommand \bibfield  [0]{\@secondoftwo}%
\providecommand \translation [1]{[#1]}%
\providecommand \BibitemOpen [0]{}%
\providecommand \bibitemStop [0]{}%
\providecommand \bibitemNoStop [0]{.\EOS\space}%
\providecommand \EOS [0]{\spacefactor3000\relax}%
\providecommand \BibitemShut  [1]{\csname bibitem#1\endcsname}%
\let\auto@bib@innerbib\@empty
%</preamble>
\bibitem [{\citenamefont {Lewis}\ and\ \citenamefont
  {Challinor}(2006)}]{Lewis:2006fu}%
  \BibitemOpen
  \bibfield  {author} {\bibinfo {author} {\bibfnamefont {Antony}\ \bibnamefont
  {Lewis}}\ and\ \bibinfo {author} {\bibfnamefont {Anthony}\ \bibnamefont
  {Challinor}},\ }\bibfield  {title} {\enquote {\bibinfo {title} {{Weak
  gravitational lensing of the CMB}},}\ }\href {\doibase
  10.1016/j.physrep.2006.03.002} {\bibfield  {journal} {\bibinfo  {journal}
  {Phys. Rept.}\ }\textbf {\bibinfo {volume} {429}},\ \bibinfo {pages} {1--65}
  (\bibinfo {year} {2006})},\ \Eprint {http://arxiv.org/abs/astro-ph/0601594}
  {arXiv:astro-ph/0601594 [astro-ph]} \BibitemShut {NoStop}%
%%CITATION = ASTRO-PH/0601594;%%
\bibitem [{\citenamefont {{Planck Collaboration VIIII}}(2018)}]{PL2018}%
  \BibitemOpen
  \bibfield  {author} {\bibinfo {author} {\bibnamefont {{Planck Collaboration
  VIIII}}} (\bibinfo {collaboration} {Planck}),\ }\bibfield  {title} {\enquote
  {\bibinfo {title} {{Planck 2018 results. VIII. Gravitational lensing}},}\
  }\href@noop {} {\  (\bibinfo {year} {2018})},\ \Eprint
  {http://arxiv.org/abs/1807.06210} {arXiv:1807.06210 [astro-ph.CO]}
  \BibitemShut {NoStop}%
%%CITATION = ARXIV:1807.06210;%%
\bibitem [{\citenamefont {Ade}\ \emph {et~al.}(2018)\citenamefont {Ade} \emph
  {et~al.}}]{Ade:2018sbj}%
  \BibitemOpen
  \bibfield  {author} {\bibinfo {author} {\bibfnamefont {Peter}\ \bibnamefont
  {Ade}} \emph {et~al.} (\bibinfo {collaboration} {Simons Observatory}),\
  }\bibfield  {title} {\enquote {\bibinfo {title} {{The Simons Observatory:
  Science goals and forecasts}},}\ }\href@noop {} {\  (\bibinfo {year}
  {2018})},\ \Eprint {http://arxiv.org/abs/1808.07445} {arXiv:1808.07445
  [astro-ph.CO]} \BibitemShut {NoStop}%
%%CITATION = ARXIV:1808.07445;%%
\bibitem [{\citenamefont {Abazajian}\ \emph {et~al.}(2016)\citenamefont
  {Abazajian} \emph {et~al.}}]{Abazajian:2016yjj}%
  \BibitemOpen
  \bibfield  {author} {\bibinfo {author} {\bibfnamefont {Kevork~N.}\
  \bibnamefont {Abazajian}} \emph {et~al.} (\bibinfo {collaboration}
  {CMB-S4}),\ }\bibfield  {title} {\enquote {\bibinfo {title} {{CMB-S4 Science
  Book, First Edition}},}\ }\href@noop {} {\  (\bibinfo {year} {2016})},\
  \Eprint {http://arxiv.org/abs/1610.02743} {arXiv:1610.02743 [astro-ph.CO]}
  \BibitemShut {NoStop}%
%%CITATION = ARXIV:1610.02743;%%
\bibitem [{\citenamefont {Sherwin}\ \emph {et~al.}(2017)\citenamefont {Sherwin}
  \emph {et~al.}}]{Sherwin:2016tyf}%
  \BibitemOpen
  \bibfield  {author} {\bibinfo {author} {\bibfnamefont {Blake~D.}\
  \bibnamefont {Sherwin}} \emph {et~al.},\ }\bibfield  {title} {\enquote
  {\bibinfo {title} {{Two-season Atacama Cosmology Telescope polarimeter
  lensing power spectrum}},}\ }\href {\doibase 10.1103/PhysRevD.95.123529}
  {\bibfield  {journal} {\bibinfo  {journal} {\prd}\ }\textbf {\bibinfo
  {volume} {95}},\ \bibinfo {pages} {123529} (\bibinfo {year} {2017})},\
  \Eprint {http://arxiv.org/abs/1611.09753} {arXiv:1611.09753 [astro-ph.CO]}
  \BibitemShut {NoStop}%
%%CITATION = ARXIV:1611.09753;%%
\bibitem [{\citenamefont {Story}\ \emph {et~al.}(2015)\citenamefont {Story}
  \emph {et~al.}}]{Story:2014hni}%
  \BibitemOpen
  \bibfield  {author} {\bibinfo {author} {\bibfnamefont {K.~T.}\ \bibnamefont
  {Story}} \emph {et~al.} (\bibinfo {collaboration} {SPT}),\ }\bibfield
  {title} {\enquote {\bibinfo {title} {{A Measurement of the Cosmic Microwave
  Background Gravitational Lensing Potential from 100 Square Degrees of SPTpol
  Data}},}\ }\href {\doibase 10.1088/0004-637X/810/1/50} {\bibfield  {journal}
  {\bibinfo  {journal} {Astrophys. J.}\ }\textbf {\bibinfo {volume} {810}},\
  \bibinfo {pages} {50} (\bibinfo {year} {2015})},\ \Eprint
  {http://arxiv.org/abs/1412.4760} {arXiv:1412.4760 [astro-ph.CO]} \BibitemShut
  {NoStop}%
%%CITATION = ARXIV:1412.4760;%%
\bibitem [{\citenamefont {Ade}\ \emph {et~al.}(2017)\citenamefont {Ade} \emph
  {et~al.}}]{Ade:2017uvt}%
  \BibitemOpen
  \bibfield  {author} {\bibinfo {author} {\bibfnamefont {P.~A.~R.}\
  \bibnamefont {Ade}} \emph {et~al.} (\bibinfo {collaboration} {POLARBEAR}),\
  }\bibfield  {title} {\enquote {\bibinfo {title} {{A Measurement of the Cosmic
  Microwave Background $B$-Mode Polarization Power Spectrum at Sub-Degree
  Scales from 2 years of POLARBEAR Data}},}\ }\href {\doibase
  10.3847/1538-4357/aa8e9f} {\bibfield  {journal} {\bibinfo  {journal}
  {Astrophys. J.}\ }\textbf {\bibinfo {volume} {848}},\ \bibinfo {pages} {121}
  (\bibinfo {year} {2017})},\ \Eprint {http://arxiv.org/abs/1705.02907}
  {arXiv:1705.02907 [astro-ph.CO]} \BibitemShut {NoStop}%
%%CITATION = ARXIV:1705.02907;%%
\bibitem [{\citenamefont {Hirata}\ and\ \citenamefont
  {Seljak}(2003{\natexlab{a}})}]{Hirata:2003ka}%
  \BibitemOpen
  \bibfield  {author} {\bibinfo {author} {\bibfnamefont {Christopher~M.}\
  \bibnamefont {Hirata}}\ and\ \bibinfo {author} {\bibfnamefont {Uros}\
  \bibnamefont {Seljak}},\ }\bibfield  {title} {\enquote {\bibinfo {title}
  {{Reconstruction of lensing from the cosmic microwave background
  polarization}},}\ }\href {\doibase 10.1103/PhysRevD.68.083002} {\bibfield
  {journal} {\bibinfo  {journal} {Phys. Rev.}\ }\textbf {\bibinfo {volume}
  {D68}},\ \bibinfo {pages} {083002} (\bibinfo {year} {2003}{\natexlab{a}})},\
  \Eprint {http://arxiv.org/abs/astro-ph/0306354} {arXiv:astro-ph/0306354
  [astro-ph]} \BibitemShut {NoStop}%
%%CITATION = ASTRO-PH/0306354;%%
\bibitem [{\citenamefont {Carron}\ and\ \citenamefont
  {Lewis}(2017)}]{Carron:2017mqf}%
  \BibitemOpen
  \bibfield  {author} {\bibinfo {author} {\bibfnamefont {Julien}\ \bibnamefont
  {Carron}}\ and\ \bibinfo {author} {\bibfnamefont {Antony}\ \bibnamefont
  {Lewis}},\ }\bibfield  {title} {\enquote {\bibinfo {title} {{Maximum a
  posteriori CMB lensing reconstruction}},}\ }\href {\doibase
  10.1103/PhysRevD.96.063510} {\bibfield  {journal} {\bibinfo  {journal}
  {\prd}\ }\textbf {\bibinfo {volume} {96}},\ \bibinfo {pages} {063510}
  (\bibinfo {year} {2017})},\ \Eprint {http://arxiv.org/abs/1704.08230}
  {arXiv:1704.08230 [astro-ph.CO]} \BibitemShut {NoStop}%
%%CITATION = ARXIV:1704.08230;%%
\bibitem [{\citenamefont {Hirata}\ and\ \citenamefont
  {Seljak}(2003{\natexlab{b}})}]{Hirata:2002jy}%
  \BibitemOpen
  \bibfield  {author} {\bibinfo {author} {\bibfnamefont {Christopher~M.}\
  \bibnamefont {Hirata}}\ and\ \bibinfo {author} {\bibfnamefont {Uros}\
  \bibnamefont {Seljak}},\ }\bibfield  {title} {\enquote {\bibinfo {title}
  {{Analyzing weak lensing of the cosmic microwave background using the
  likelihood function}},}\ }\href {\doibase 10.1103/PhysRevD.67.043001}
  {\bibfield  {journal} {\bibinfo  {journal} {Phys. Rev.}\ }\textbf {\bibinfo
  {volume} {D67}},\ \bibinfo {pages} {043001} (\bibinfo {year}
  {2003}{\natexlab{b}})},\ \Eprint {http://arxiv.org/abs/astro-ph/0209489}
  {arXiv:astro-ph/0209489 [astro-ph]} \BibitemShut {NoStop}%
%%CITATION = ASTRO-PH/0209489;%%
\bibitem [{\citenamefont {Okamoto}\ and\ \citenamefont
  {Hu}(2003)}]{Okamoto:2003zw}%
  \BibitemOpen
  \bibfield  {author} {\bibinfo {author} {\bibfnamefont {Takemi}\ \bibnamefont
  {Okamoto}}\ and\ \bibinfo {author} {\bibfnamefont {Wayne}\ \bibnamefont
  {Hu}},\ }\bibfield  {title} {\enquote {\bibinfo {title} {{CMB lensing
  reconstruction on the full sky}},}\ }\href {\doibase
  10.1103/PhysRevD.67.083002} {\bibfield  {journal} {\bibinfo  {journal} {Phys.
  Rev.}\ }\textbf {\bibinfo {volume} {D67}},\ \bibinfo {pages} {083002}
  (\bibinfo {year} {2003})},\ \Eprint {http://arxiv.org/abs/astro-ph/0301031}
  {arXiv:astro-ph/0301031 [astro-ph]} \BibitemShut {NoStop}%
%%CITATION = ASTRO-PH/0301031;%%
\bibitem [{\citenamefont {Hanson}\ \emph {et~al.}(2010)\citenamefont {Hanson},
  \citenamefont {Challinor},\ and\ \citenamefont {Lewis}}]{Hanson:2009kr}%
  \BibitemOpen
  \bibfield  {author} {\bibinfo {author} {\bibfnamefont {Duncan}\ \bibnamefont
  {Hanson}}, \bibinfo {author} {\bibfnamefont {Anthony}\ \bibnamefont
  {Challinor}}, \ and\ \bibinfo {author} {\bibfnamefont {Antony}\ \bibnamefont
  {Lewis}},\ }\bibfield  {title} {\enquote {\bibinfo {title} {{Weak lensing of
  the CMB}},}\ }\href {\doibase 10.1007/s10714-010-1036-y} {\bibfield
  {journal} {\bibinfo  {journal} {Gen. Rel. Grav.}\ }\textbf {\bibinfo {volume}
  {42}},\ \bibinfo {pages} {2197--2218} (\bibinfo {year} {2010})},\ \Eprint
  {http://arxiv.org/abs/0911.0612} {arXiv:0911.0612 [astro-ph.CO]} \BibitemShut
  {NoStop}%
%%CITATION = ARXIV:0911.0612;%%
\bibitem [{\citenamefont {{Planck Collaboration XIII}}(2016)}]{Ade:2015xua}%
  \BibitemOpen
  \bibfield  {author} {\bibinfo {author} {\bibnamefont {{Planck Collaboration
  XIII}}} (\bibinfo {collaboration} {Planck}),\ }\bibfield  {title} {\enquote
  {\bibinfo {title} {{Planck 2015 results. XIII. Cosmological parameters}},}\
  }\href {\doibase 10.1051/0004-6361/201525830} {\bibfield  {journal} {\bibinfo
   {journal} {\aap}\ }\textbf {\bibinfo {volume} {594}},\ \bibinfo {pages}
  {A13} (\bibinfo {year} {2016})},\ \Eprint {http://arxiv.org/abs/1502.01589}
  {arXiv:1502.01589 [astro-ph.CO]} \BibitemShut {NoStop}%
%%CITATION = ARXIV:1502.01589;%%
\bibitem [{\citenamefont {Hanson}\ \emph {et~al.}(2011)\citenamefont {Hanson},
  \citenamefont {Challinor}, \citenamefont {Efstathiou},\ and\ \citenamefont
  {Bielewicz}}]{Hanson:2010rp}%
  \BibitemOpen
  \bibfield  {author} {\bibinfo {author} {\bibfnamefont {Duncan}\ \bibnamefont
  {Hanson}}, \bibinfo {author} {\bibfnamefont {Anthony}\ \bibnamefont
  {Challinor}}, \bibinfo {author} {\bibfnamefont {George}\ \bibnamefont
  {Efstathiou}}, \ and\ \bibinfo {author} {\bibfnamefont {Pawel}\ \bibnamefont
  {Bielewicz}},\ }\bibfield  {title} {\enquote {\bibinfo {title} {{CMB
  temperature lensing power reconstruction}},}\ }\href@noop {} {\bibfield
  {journal} {\bibinfo  {journal} {\prd}\ }\textbf {\bibinfo {volume} {83}},\
  \bibinfo {pages} {043005} (\bibinfo {year} {2011})},\ \Eprint
  {http://arxiv.org/abs/1008.4403} {arXiv:1008.4403 [astro-ph.CO]} \BibitemShut
  {NoStop}%
%%CITATION = 1008.4403;%%
\bibitem [{\citenamefont {Lewis}\ \emph {et~al.}(2011)\citenamefont {Lewis},
  \citenamefont {Challinor},\ and\ \citenamefont {Hanson}}]{Lewis:2011fk}%
  \BibitemOpen
  \bibfield  {author} {\bibinfo {author} {\bibfnamefont {Antony}\ \bibnamefont
  {Lewis}}, \bibinfo {author} {\bibfnamefont {Anthony}\ \bibnamefont
  {Challinor}}, \ and\ \bibinfo {author} {\bibfnamefont {Duncan}\ \bibnamefont
  {Hanson}},\ }\bibfield  {title} {\enquote {\bibinfo {title} {{The shape of
  the CMB lensing bispectrum}},}\ }\href {\doibase
  10.1088/1475-7516/2011/03/018} {\bibfield  {journal} {\bibinfo  {journal}
  {JCAP}\ }\textbf {\bibinfo {volume} {1103}},\ \bibinfo {pages} {018}
  (\bibinfo {year} {2011})},\ \Eprint {http://arxiv.org/abs/1101.2234}
  {arXiv:1101.2234 [astro-ph.CO]} \BibitemShut {NoStop}%
%%CITATION = ARXIV:1101.2234;%%
\bibitem [{\citenamefont {Schmittfull}\ \emph {et~al.}(2013)\citenamefont
  {Schmittfull}, \citenamefont {Challinor}, \citenamefont {Hanson},\ and\
  \citenamefont {Lewis}}]{Schmittfull:2013uea}%
  \BibitemOpen
  \bibfield  {author} {\bibinfo {author} {\bibfnamefont {Marcel~M.}\
  \bibnamefont {Schmittfull}}, \bibinfo {author} {\bibfnamefont {Anthony}\
  \bibnamefont {Challinor}}, \bibinfo {author} {\bibfnamefont {Duncan}\
  \bibnamefont {Hanson}}, \ and\ \bibinfo {author} {\bibfnamefont {Antony}\
  \bibnamefont {Lewis}},\ }\bibfield  {title} {\enquote {\bibinfo {title} {{On
  the joint analysis of CMB temperature and lensing-reconstruction power
  spectra}},}\ }\href {\doibase 10.1103/PhysRevD.88.063012} {\bibfield
  {journal} {\bibinfo  {journal} {\prd}\ }\textbf {\bibinfo {volume} {88}},\
  \bibinfo {pages} {063012} (\bibinfo {year} {2013})},\ \Eprint
  {http://arxiv.org/abs/1308.0286} {arXiv:1308.0286 [astro-ph.CO]} \BibitemShut
  {NoStop}%
%%CITATION = ARXIV:1308.0286;%%
\bibitem [{\citenamefont {Stevens}\ \emph {et~al.}(2018)\citenamefont {Stevens}
  \emph {et~al.}}]{Stevens:2018biw}%
  \BibitemOpen
  \bibfield  {author} {\bibinfo {author} {\bibfnamefont {Jason~R.}\
  \bibnamefont {Stevens}} \emph {et~al.},\ }\bibfield  {title} {\enquote
  {\bibinfo {title} {{Designs for next generation CMB survey strategies from
  Chile}},}\ }\href {\doibase 10.1117/12.2313898} {\bibfield  {journal}
  {\bibinfo  {journal} {Proc. SPIE Int. Soc. Opt. Eng.}\ }\textbf {\bibinfo
  {volume} {10708}},\ \bibinfo {pages} {1070841} (\bibinfo {year} {2018})},\
  \Eprint {http://arxiv.org/abs/1808.05131} {arXiv:1808.05131 [astro-ph.IM]}
  \BibitemShut {NoStop}%
%%CITATION = ARXIV:1808.05131;%%
\bibitem [{\citenamefont {Smith}\ \emph {et~al.}(2007)\citenamefont {Smith},
  \citenamefont {Zahn},\ and\ \citenamefont {Dore}}]{Smith:2007rg}%
  \BibitemOpen
  \bibfield  {author} {\bibinfo {author} {\bibfnamefont {Kendrick~M.}\
  \bibnamefont {Smith}}, \bibinfo {author} {\bibfnamefont {Oliver}\
  \bibnamefont {Zahn}}, \ and\ \bibinfo {author} {\bibfnamefont {Olivier}\
  \bibnamefont {Dore}},\ }\bibfield  {title} {\enquote {\bibinfo {title}
  {{Detection of Gravitational Lensing in the Cosmic Microwave Background}},}\
  }\href {\doibase 10.1103/PhysRevD.76.043510} {\bibfield  {journal} {\bibinfo
  {journal} {Phys. Rev.}\ }\textbf {\bibinfo {volume} {D76}},\ \bibinfo {pages}
  {043510} (\bibinfo {year} {2007})},\ \Eprint {http://arxiv.org/abs/0705.3980}
  {arXiv:0705.3980 [astro-ph]} \BibitemShut {NoStop}%
%%CITATION = ARXIV:0705.3980;%%
\bibitem [{\citenamefont {Ade}\ \emph {et~al.}(2014)\citenamefont {Ade} \emph
  {et~al.}}]{Ade:2013tyw}%
  \BibitemOpen
  \bibfield  {author} {\bibinfo {author} {\bibfnamefont {P.~A.~R.}\
  \bibnamefont {Ade}} \emph {et~al.} (\bibinfo {collaboration} {Planck}),\
  }\bibfield  {title} {\enquote {\bibinfo {title} {{Planck 2013 results. XVII.
  Gravitational lensing by large-scale structure}},}\ }\href {\doibase
  10.1051/0004-6361/201321543} {\bibfield  {journal} {\bibinfo  {journal}
  {Astron. Astrophys.}\ }\textbf {\bibinfo {volume} {571}},\ \bibinfo {pages}
  {A17} (\bibinfo {year} {2014})},\ \Eprint {http://arxiv.org/abs/1303.5077}
  {arXiv:1303.5077 [astro-ph.CO]} \BibitemShut {NoStop}%
%%CITATION = ARXIV:1303.5077;%%
\bibitem [{\citenamefont {Ade}\ \emph {et~al.}(2016{\natexlab{a}})\citenamefont
  {Ade} \emph {et~al.}}]{Ade:2015zua}%
  \BibitemOpen
  \bibfield  {author} {\bibinfo {author} {\bibfnamefont {P.~A.~R.}\
  \bibnamefont {Ade}} \emph {et~al.} (\bibinfo {collaboration} {Planck}),\
  }\bibfield  {title} {\enquote {\bibinfo {title} {{Planck 2015 results. XV.
  Gravitational lensing}},}\ }\href {\doibase 10.1051/0004-6361/201525941}
  {\bibfield  {journal} {\bibinfo  {journal} {Astron. Astrophys.}\ }\textbf
  {\bibinfo {volume} {594}},\ \bibinfo {pages} {A15} (\bibinfo {year}
  {2016}{\natexlab{a}})},\ \Eprint {http://arxiv.org/abs/1502.01591}
  {arXiv:1502.01591 [astro-ph.CO]} \BibitemShut {NoStop}%
%%CITATION = ARXIV:1502.01591;%%
\bibitem [{\citenamefont {Benoit-Levy}\ \emph {et~al.}(2013)\citenamefont
  {Benoit-Levy}, \citenamefont {Dechelette}, \citenamefont {Benabed},
  \citenamefont {Cardoso}, \citenamefont {Hanson} \emph
  {et~al.}}]{BenoitLevy:2013bc}%
  \BibitemOpen
  \bibfield  {author} {\bibinfo {author} {\bibfnamefont {Aurelien}\
  \bibnamefont {Benoit-Levy}}, \bibinfo {author} {\bibfnamefont {Typhaine}\
  \bibnamefont {Dechelette}}, \bibinfo {author} {\bibfnamefont {Karim}\
  \bibnamefont {Benabed}}, \bibinfo {author} {\bibfnamefont {Jean-Francois}\
  \bibnamefont {Cardoso}}, \bibinfo {author} {\bibfnamefont {Duncan}\
  \bibnamefont {Hanson}},  \emph {et~al.},\ }\bibfield  {title} {\enquote
  {\bibinfo {title} {{Full-sky CMB lensing reconstruction in presence of
  sky-cuts}},}\ }\href {\doibase 10.1051/0004-6361/201321048} {\bibfield
  {journal} {\bibinfo  {journal} {\aap}\ }\textbf {\bibinfo {volume} {555}},\
  \bibinfo {pages} {A37} (\bibinfo {year} {2013})},\ \Eprint
  {http://arxiv.org/abs/1301.4145} {arXiv:1301.4145 [astro-ph.CO]} \BibitemShut
  {NoStop}%
%%CITATION = ARXIV:1301.4145;%%
\bibitem [{\citenamefont {Smith}(2006)}]{Smith:2005gi}%
  \BibitemOpen
  \bibfield  {author} {\bibinfo {author} {\bibfnamefont {Kendrick~M.}\
  \bibnamefont {Smith}},\ }\bibfield  {title} {\enquote {\bibinfo {title}
  {{Pseudo-$C_\ell$ estimators which do not mix E and B modes}},}\ }\href
  {\doibase 10.1103/PhysRevD.74.083002} {\bibfield  {journal} {\bibinfo
  {journal} {Phys. Rev.}\ }\textbf {\bibinfo {volume} {D74}},\ \bibinfo {pages}
  {083002} (\bibinfo {year} {2006})},\ \Eprint
  {http://arxiv.org/abs/astro-ph/0511629} {arXiv:astro-ph/0511629 [astro-ph]}
  \BibitemShut {NoStop}%
%%CITATION = ASTRO-PH/0511629;%%
\bibitem [{\citenamefont {Pearson}\ \emph {et~al.}(2014)\citenamefont
  {Pearson}, \citenamefont {Sherwin},\ and\ \citenamefont
  {Lewis}}]{Pearson:2014qna}%
  \BibitemOpen
  \bibfield  {author} {\bibinfo {author} {\bibfnamefont {Ruth}\ \bibnamefont
  {Pearson}}, \bibinfo {author} {\bibfnamefont {Blake}\ \bibnamefont
  {Sherwin}}, \ and\ \bibinfo {author} {\bibfnamefont {Antony}\ \bibnamefont
  {Lewis}},\ }\bibfield  {title} {\enquote {\bibinfo {title} {{CMB lensing
  reconstruction using cut sky polarization maps and pure-$B$ modes}},}\ }\href
  {\doibase 10.1103/PhysRevD.90.023539} {\bibfield  {journal} {\bibinfo
  {journal} {Phys. Rev.}\ }\textbf {\bibinfo {volume} {D90}},\ \bibinfo {pages}
  {023539} (\bibinfo {year} {2014})},\ \Eprint {http://arxiv.org/abs/1403.3911}
  {arXiv:1403.3911 [astro-ph.CO]} \BibitemShut {NoStop}%
%%CITATION = ARXIV:1403.3911;%%
\bibitem [{\citenamefont {Ade}\ \emph {et~al.}(2016{\natexlab{b}})\citenamefont
  {Ade} \emph {et~al.}}]{Array:2016afx}%
  \BibitemOpen
  \bibfield  {author} {\bibinfo {author} {\bibfnamefont {P.~A.~R.}\
  \bibnamefont {Ade}} \emph {et~al.} (\bibinfo {collaboration} {BICEP2, Keck
  Array}),\ }\bibfield  {title} {\enquote {\bibinfo {title} {{BICEP2 / Keck
  Array VIII: Measurement of gravitational lensing from large-scale B-mode
  polarization}},}\ }\href {\doibase 10.3847/1538-4357/833/2/228} {\bibfield
  {journal} {\bibinfo  {journal} {Astrophys. J.}\ }\textbf {\bibinfo {volume}
  {833}},\ \bibinfo {pages} {228} (\bibinfo {year} {2016}{\natexlab{b}})},\
  \Eprint {http://arxiv.org/abs/1606.01968} {arXiv:1606.01968 [astro-ph.CO]}
  \BibitemShut {NoStop}%
%%CITATION = ARXIV:1606.01968;%%
\bibitem [{\citenamefont {Kesden}\ \emph {et~al.}(2003)\citenamefont {Kesden},
  \citenamefont {Cooray},\ and\ \citenamefont {Kamionkowski}}]{Kesden:2003cc}%
  \BibitemOpen
  \bibfield  {author} {\bibinfo {author} {\bibfnamefont {Michael}\ \bibnamefont
  {Kesden}}, \bibinfo {author} {\bibfnamefont {Asantha}\ \bibnamefont
  {Cooray}}, \ and\ \bibinfo {author} {\bibfnamefont {Marc}\ \bibnamefont
  {Kamionkowski}},\ }\bibfield  {title} {\enquote {\bibinfo {title} {Lensing
  reconstruction with {CMB} temperature and polarization},}\ }\href@noop {}
  {\bibfield  {journal} {\bibinfo  {journal} {\prd}\ }\textbf {\bibinfo
  {volume} {67}},\ \bibinfo {pages} {123507} (\bibinfo {year} {2003})},\
  \Eprint {http://arxiv.org/abs/astro-ph/0302536} {astro-ph/0302536}
  \BibitemShut {NoStop}%
%%CITATION = ASTRO-PH 0302536;%%
\bibitem [{\citenamefont {Hu}\ and\ \citenamefont {Okamoto}(2002)}]{Hu:2001kj}%
  \BibitemOpen
  \bibfield  {author} {\bibinfo {author} {\bibfnamefont {Wayne}\ \bibnamefont
  {Hu}}\ and\ \bibinfo {author} {\bibfnamefont {Takemi}\ \bibnamefont
  {Okamoto}},\ }\bibfield  {title} {\enquote {\bibinfo {title} {{Mass
  reconstruction with CMB polarization}},}\ }\href {\doibase 10.1086/341110}
  {\bibfield  {journal} {\bibinfo  {journal} {Astrophys. J.}\ }\textbf
  {\bibinfo {volume} {574}},\ \bibinfo {pages} {566--574} (\bibinfo {year}
  {2002})},\ \Eprint {http://arxiv.org/abs/astro-ph/0111606}
  {arXiv:astro-ph/0111606 [astro-ph]} \BibitemShut {NoStop}%
%%CITATION = ASTRO-PH/0111606;%%
\bibitem [{\citenamefont {Fabbian}\ \emph {et~al.}(2019)\citenamefont
  {Fabbian}, \citenamefont {Lewis},\ and\ \citenamefont
  {Beck}}]{Fabbian:2019tik}%
  \BibitemOpen
  \bibfield  {author} {\bibinfo {author} {\bibfnamefont {Giulio}\ \bibnamefont
  {Fabbian}}, \bibinfo {author} {\bibfnamefont {Antony}\ \bibnamefont {Lewis}},
  \ and\ \bibinfo {author} {\bibfnamefont {Dominic}\ \bibnamefont {Beck}},\
  }\bibfield  {title} {\enquote {\bibinfo {title} {{CMB lensing reconstruction
  biases in cross-correlation with large-scale structure probes}},}\
  }\href@noop {} {\  (\bibinfo {year} {2019})},\ \Eprint
  {http://arxiv.org/abs/1906.08760} {arXiv:1906.08760 [astro-ph.CO]}
  \BibitemShut {NoStop}%
%%CITATION = ARXIV:1906.08760;%%
\bibitem [{\citenamefont {Lewis}\ \emph {et~al.}(2000)\citenamefont {Lewis},
  \citenamefont {Challinor},\ and\ \citenamefont {Lasenby}}]{Lewis:1999bs}%
  \BibitemOpen
  \bibfield  {author} {\bibinfo {author} {\bibfnamefont {Antony}\ \bibnamefont
  {Lewis}}, \bibinfo {author} {\bibfnamefont {Anthony}\ \bibnamefont
  {Challinor}}, \ and\ \bibinfo {author} {\bibfnamefont {Anthony}\ \bibnamefont
  {Lasenby}},\ }\bibfield  {title} {\enquote {\bibinfo {title} {{Efficient
  computation of CMB anisotropies in closed FRW models}},}\ }\href {\doibase
  10.1086/309179} {\bibfield  {journal} {\bibinfo  {journal} {\apj}\ }\textbf
  {\bibinfo {volume} {538}},\ \bibinfo {pages} {473--476} (\bibinfo {year}
  {2000})},\ \Eprint {http://arxiv.org/abs/astro-ph/9911177}
  {arXiv:astro-ph/9911177 [astro-ph]} \BibitemShut {NoStop}%
%%CITATION = ASTRO-PH/9911177;%%
\bibitem [{\citenamefont {Plaszczynski}\ \emph {et~al.}(2012)\citenamefont
  {Plaszczynski}, \citenamefont {Lavabre}, \citenamefont {Perotto},\ and\
  \citenamefont {Starck}}]{Plaszczynski:2012ej}%
  \BibitemOpen
  \bibfield  {author} {\bibinfo {author} {\bibfnamefont {S.}~\bibnamefont
  {Plaszczynski}}, \bibinfo {author} {\bibfnamefont {A.}~\bibnamefont
  {Lavabre}}, \bibinfo {author} {\bibfnamefont {L.}~\bibnamefont {Perotto}}, \
  and\ \bibinfo {author} {\bibfnamefont {J-L}\ \bibnamefont {Starck}},\
  }\bibfield  {title} {\enquote {\bibinfo {title} {{A hybrid approach to CMB
  lensing reconstruction on all-sky intensity maps}},}\ }\href {\doibase
  10.1051/0004-6361/201218899} {\bibfield  {journal} {\bibinfo  {journal}
  {Astron. Astrophys.}\ }\textbf {\bibinfo {volume} {544}},\ \bibinfo {pages}
  {A27} (\bibinfo {year} {2012})},\ \Eprint {http://arxiv.org/abs/1201.5779}
  {arXiv:1201.5779 [astro-ph.CO]} \BibitemShut {NoStop}%
%%CITATION = ARXIV:1201.5779;%%
\bibitem [{\citenamefont {Namikawa}\ and\ \citenamefont
  {Nagata}(2014)}]{Namikawa:2014yca}%
  \BibitemOpen
  \bibfield  {author} {\bibinfo {author} {\bibfnamefont {Toshiya}\ \bibnamefont
  {Namikawa}}\ and\ \bibinfo {author} {\bibfnamefont {Ryo}\ \bibnamefont
  {Nagata}},\ }\bibfield  {title} {\enquote {\bibinfo {title} {{Lensing
  reconstruction from a patchwork of polarization maps}},}\ }\href {\doibase
  10.1088/1475-7516/2014/09/009} {\bibfield  {journal} {\bibinfo  {journal}
  {\jcap}\ }\textbf {\bibinfo {volume} {1409}},\ \bibinfo {pages} {009}
  (\bibinfo {year} {2014})},\ \Eprint {http://arxiv.org/abs/1405.6568}
  {arXiv:1405.6568 [astro-ph.CO]} \BibitemShut {NoStop}%
%%CITATION = ARXIV:1405.6568;%%
\bibitem [{\citenamefont {{Abazajian}}\ \emph {et~al.}(2019)\citenamefont
  {{Abazajian}} \emph {et~al.}}]{2019arXiv190801062A}%
  \BibitemOpen
  \bibfield  {author} {\bibinfo {author} {\bibfnamefont {Kevork}\ \bibnamefont
  {{Abazajian}}} \emph {et~al.},\ }\bibfield  {title} {\enquote {\bibinfo
  {title} {{CMB-S4 Decadal Survey APC White Paper}},}\ }\href@noop {} {\
  (\bibinfo {year} {2019})},\ \Eprint {http://arxiv.org/abs/1908.01062}
  {arXiv:1908.01062 [astro-ph.IM]} \BibitemShut {NoStop}%
\bibitem [{\citenamefont {Münchmeyer}\ and\ \citenamefont
  {Smith}(2019)}]{Munchmeyer:2019kng}%
  \BibitemOpen
  \bibfield  {author} {\bibinfo {author} {\bibfnamefont {Moritz}\ \bibnamefont
  {Münchmeyer}}\ and\ \bibinfo {author} {\bibfnamefont {Kendrick~M.}\
  \bibnamefont {Smith}},\ }\bibfield  {title} {\enquote {\bibinfo {title}
  {{Fast Wiener filtering of CMB maps with Neural Networks}},}\ }\href@noop {}
  {\  (\bibinfo {year} {2019})},\ \Eprint {http://arxiv.org/abs/1905.05846}
  {arXiv:1905.05846 [astro-ph.CO]} \BibitemShut {NoStop}%
%%CITATION = ARXIV:1905.05846;%%
\bibitem [{\citenamefont {Gorski}\ \emph {et~al.}(2005)\citenamefont {Gorski}
  \emph {et~al.}}]{Gorski:2004by}%
  \BibitemOpen
  \bibfield  {author} {\bibinfo {author} {\bibfnamefont {K.~M.}\ \bibnamefont
  {Gorski}} \emph {et~al.},\ }\bibfield  {title} {\enquote {\bibinfo {title}
  {Healpix -- a framework for high resolution discretization, and fast analysis
  of data distributed on the sphere},}\ }\href@noop {} {\bibfield  {journal}
  {\bibinfo  {journal} {\apj}\ }\textbf {\bibinfo {volume} {622}},\ \bibinfo
  {pages} {759--771} (\bibinfo {year} {2005})},\ \Eprint
  {http://arxiv.org/abs/astro-ph/0409513} {astro-ph/0409513} \BibitemShut
  {NoStop}%
%%CITATION = ASTRO-PH 0409513;%%
\end{thebibliography}%

%\clearpage

%%%%%%%%%%%%%%%%%%%%%%%%%%%%%%%%%%%%%%%%%%%%
%%%%%%%%%%%%%%%%%%%%%%%%%%%%%%%%%%%%%%%%%%%%
\appendix

\section{Optimal filtering performance}
\label{app:cgconvergence}

\begin{figure*}
\includegraphics[width=\textwidth]{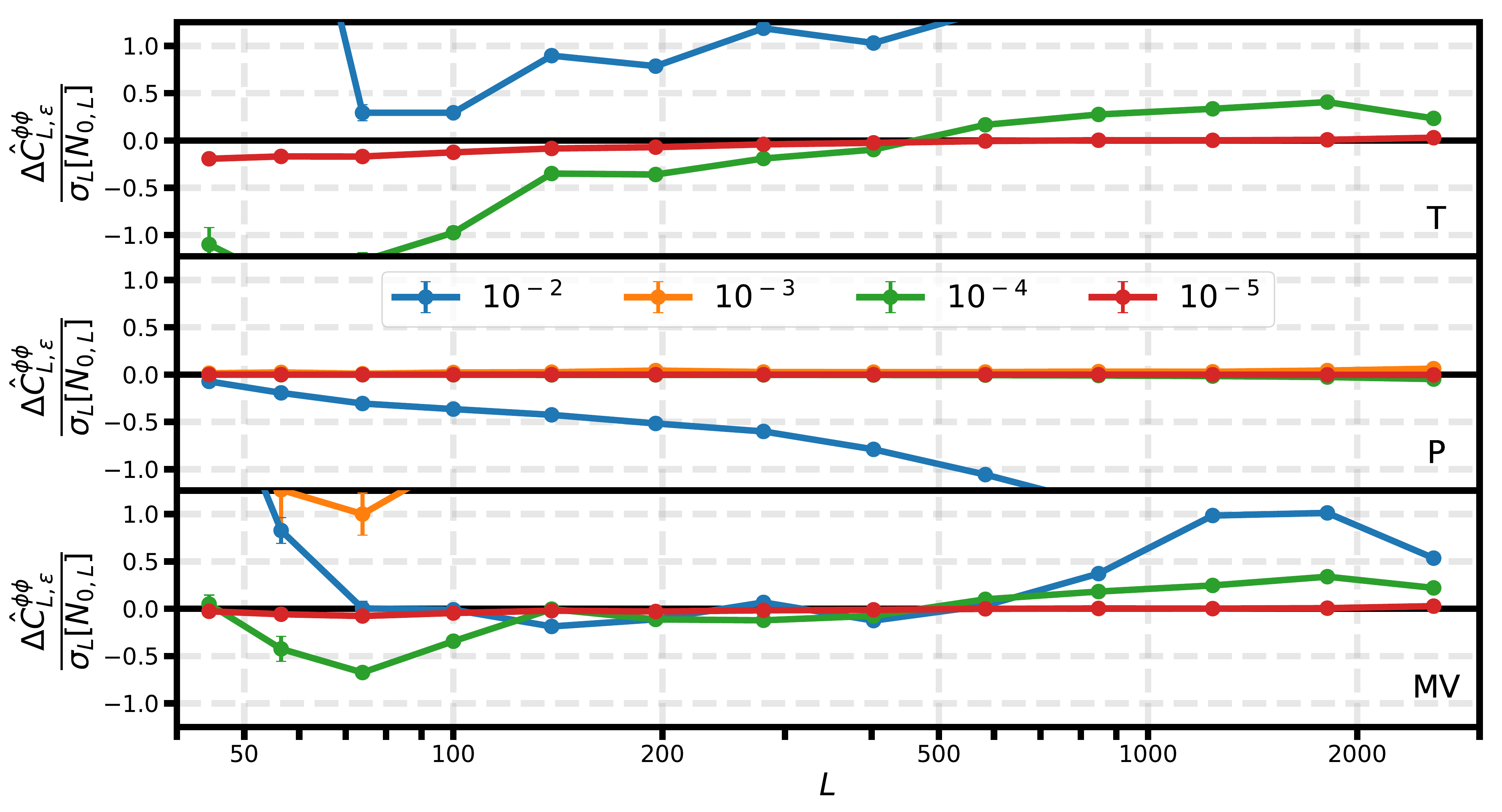}
\caption{Lensing power spectra differences built from partially conjugate-gradient-converged inverse-variance filtered CMB maps, in units of the reconstruction noise statistical error. The points shown are for the T (top panel), P (middle panel) or MV (bottom panel) lensing spectrum estimates, for convergence levels $\epsilon = 10^{-2}$ (blue), $10^{-3}$ (orange), $10^{-4}$ (green), $10^{-5}$ (red), where $\epsilon$ monitors the decrease of the residual vector norm. Convergence for P is always substantially faster than for T or MV. The points with $\epsilon=10^{-3}$ are off the scale for T and most of MV, but close to the zero for P. Our baseline T and MV analyses use $\epsilon = 10^{-5}$, where the bias is at most $0.2\sigma[ N_{0}]$ in T, and $\epsilon = 10^{-4}$ for P analysis. For T and MV the convergence is not monotonic; small changes in the high-$\ell$ temperature maps in the iterative search have a large impact on the lensing power iterations.
}
\label{fig:cgconvergence}
\end{figure*}

Optimal filtering of the CMB maps (Eq.~\eqref{eq:optfilt}) must be performed iteratively to have a tractable numerical cost. This appendix discusses the performance of our solver and convergence criteria, for both temperature and polarization.

We use the same algorithms used for the Planck lensing 2015~\citep{Ade:2015zua} and 2018~\citep{PL2018} analyses, using the flat-sky implementation in the \lensit package. The inversion is performed using a simple conjugate-gradient descent, which can be coupled to a multi-grid preconditioner~\citep{Smith:2007rg} to try and improve the convergence properties of the iterative search. When solving for $\bf{x}$ in the linear equation $\bf{x} = \bf{A}^{-1} \bf{b}$, we determine convergence by monitoring the ratio $\epsilon_n^2$ of the squared norm of the residual vector at iteration $n$, $\bf{A} \bf{x_n} - \bf{b}$, to that of the initial residual $\bf{b}$ (we always start with $\bf{x_0} = 0$). We then ensure that this ratio has dropped below a specific tolerance level $\epsilon$.

Fig.~\ref{fig:cgconvergence} shows the dependence of the lensing reconstruction on the choice of tolerance level, in units of the statistical error bars, when using the simplest possible diagonal isotropic preconditioner.
The figure shows the difference to the result for $\epsilon = 10^{-6}$ (the smallest tolerance we consider) in units of the expected statistical error bars $\sigma_L[ N_{0}]$ induced by the reconstruction noise, but excluding the lensing map cosmic variance. For this we use the approximation
\begin{equation}\label{eq:s2LN0}
\sigma_L^{2}[ N_{0,L} ]
 \equiv \frac{2 \left( \MCN_{0,L}\right)^2}{\fpatch n_L \Delta L}.
\end{equation}
These noise-only error bars are the most relevant for applications where the lensing map rather than its spectrum must be reconstructed accurately (such as for delensing).
Fig.~\ref{fig:cgconvergence} shows the power spectrum difference at the iteration for which $\epsilon \simeq 10^{-2}$ (blue), $10^{-3}$ (orange), $10^{-4}$ (green), $10^{-5}$ (red), and for T, P and MV analyses (from top to bottom). Polarization filtering is a lot better behaved than temperature. This is striking, both from the figure and in the number of iterations required to reach these tolerance levels: about $80$ steps are necessary to reach $\epsilon = 10^{-4}$ for temperature or MV filtering, but $50$ for polarization. For $\epsilon = 10^{-5}$ this becomes 430 and 100, and for $\epsilon = 10^{-6}$ this worsens to 2480 and 125, respectively.

Our T and MV baseline results used a tolerance of $10^{-5}$ and our P results used $10^{-4}$, which is enough to ensure convergence on all relevant scales to below $0.2\sigma[N_{0,L}]$. To test for broadband correlations we have evaluated biases in the overall reconstructed lensing spectrum amplitude in the following way. Using an approximation to the spectrum variance $\sigma_L^{2}[C^{\phi\phi} + N_{0}]$ (defined as in Eq.~\ref{eq:s2LN0}, but including the lens cosmic variance), we test for a non-zero lensing amplitude in the residual
\begin{equation}
\hat C_{L, \epsilon}^{\phi\phi} - \hat C_{L, \epsilon = 10^{-6}}^{\phi\phi} \equiv A_{\epsilon} C_L^{\phi\phi, \fid}
\end{equation}
using the inverse-variance weighting amplitude estimator
\begin{equation}
	\hat A_{\epsilon} = \sigma^2_{A}\sum_{L = 40}^{3000} \frac{C_L^{\phi\phi, \fid}\left(\hat C_{L, \epsilon}^{\phi\phi} - \hat C_{L,\epsilon= 10^{-6}}^{\phi\phi}\right)}{\sigma_L^{2}[C_L^{\phi\phi, \fid} + N_{0,L}]},
\end{equation}
with normalization
\begin{equation}
	\frac{1}{\sigma^2_{A}} = \sum_{L = 40}^{3000} \frac{\left(C^{\phi\phi, \fid}_L \right)^2}{\sigma_L^{2}[C_L^{\phi\phi, \fid} + N_{0,L}]}.
\end{equation}
$\sigma^2_{A}$ is also the Fisher variance $\var{\hat A_\epsilon}$ of the lensing amplitude estimator.
For $\epsilon = 10^{-5}$, in temperature case we find a maximal bias of $0.06\sigma_A$. For the cases of polarization and MV, the bias is much smaller still. It takes only about $5$ minutes on a standard laptop to filter the maps in temperature or polarization, and $15$ for joint filtering, hence the entire process remains very practical. We use the same filtering method to filter the reconstructed $\kappa$ map. In this case, the convergence is much faster, $\lesssim1$ minute for $\epsilon = 10^{-7}$, mostly because the noise is much larger compared to the lensing signal.

Our optimal CMB filtering step remains fast and easily fast enough for use in many simulations. However, we used the flat-sky approximation throughout, where harmonic transforms are fast Fourier transforms, which will not be adequate for future surveys covering a significant fraction of the sky. On the curved sky the filtering step will be more time consuming given the much more expensive cost of the high-resolution spherical harmonic transforms, and further numerical optimization of the filter will potentially be of significant benefit. For our flat sky analysis we investigated several multigrid preconditioners but found no clear-cut improvements in execution time. However, this conclusion may well change on the curved sky where the numerical cost is distributed differently. Given the very configuration-specific timing performance of the inversion, it seems likely that a good solution is probably best found on a case-by-case basis. For these reasons we do not perform further performance comparisons here and leave a systematic study of the conjugate-gradient inversion performance for future work. The convergence criteria could also be refined to more closely match the actual requirements in terms of numerical precision of the filtered maps.

\end{document}